\documentclass[12pt]{article}
\usepackage[super,compress]{cite}
\usepackage{graphicx}
\usepackage{epsfig}
\usepackage{amssymb}
\begin{document}

\begin{center}
{\bf HEAVY QUARK PRODUCTION IN $K_T$ FACTORIZATION
APPROACH AT LHC ENERGIES
}

\vspace{.2cm}

Yu.M. Shabelski$^1$, A.G. Shuvaev$^1$,
I.V. Surnin$^{1,2}$ \\

\vspace{.5cm}

$^1$Petersburg Nuclear Physics Institute, Kurchatov National
Research Centre\\
Gatchina, St. Petersburg 188300, Russia\\
$^2$Saint Petersburg State University, St.-Petersburg, Russia

\vskip 0.9 truecm
E-mail: shabelsk@thd.pnpi.spb.ru\\
E-mail: shuvaev@thd.pnpi.spb.ru\\
E-mail: surnin.ivan@mail.ru

\end{center}

\vspace{1.2cm}

\begin{abstract}
A new version of the $k_T$ factorization approach
is formulated for the high energy heavy quark production.
The results are in reasonable agreement with the experimental
data at LHC energies.
\end{abstract}

\section{Introduction}
The description of hard interactions in hadron collisions
is carried out in the perturbative QCD on the basis of
parton distribution functions. The hard cross section
results from the convolution of the incident
partons' densities with the squared sub-process amplitude
of their scattering. While some phenomenology is needed
to find the first the latter is evaluated perturbatively.
The simplest and most popular way to do it
is the parton model~\cite{PM1, PM2, PM3}.

The parton model rests upon the collinear
approximation, according to which the partons
participating in the scattering
stem from the subsequent emission off the colliding
hadrons.
The angles at which they are emitted,
or their transverse momenta,
increase for each consecutive emission
reaching the top value for the partons involved in
the hard subprocess. The evolution of the parton
distribution as a function
of maximal allowed transverse momentum square $Q^2$
is governed by DGLAP equation that collect
the large terms $\log Q^2/\mu^2$ for a certain scale
$\mu^2$~\cite{DDT}.

The value $Q^2$ is supposed to be negligible
compared to the transverse momenta of
the heavy quarks in the conventional
parton model.
As a virtuality of the emitted partons
is of the order of their transverse momenta
in the leading logarithmic approximation
they are treated as mass shell particles
with purely longitudinal momenta.
The heavy quarks are produced therefore
back to back so that the total transverse
momentum of the quark-antiquark pair is always zero.
The virtuality of the incoming partons
is taken into account only
through $Q^2$ dependence of the structure functions.
A more elaborated
kinematics, in particular non vanishing pair
transverse momentum, requires to go beyond
the leading order (LO) of the parton model.

There is another parameter that becomes significant
for the very high energy, $\log 1/x$, where $x$ is
the total momentum fraction carried by a parton.
The small $x$ is the region where BFKL dynamics
works \cite{BFKL1, BFKL2}.
An effective approach to the dynamics for $x \ll 1$
and large $Q^2$ is $k_T$ factorization
method~\cite{kT1, kT2, kT3, kT4}, in which
the partons are assumed to be virtual like in
the Feynman diagrams.
The basic difference from the conventional parton model
is that the partons in this approach poses
an intrinsic transfer momentum, which is not more neglected.
The leading order of the $k_T$
factorization embodies not only LO of the parton model
but an essential part of the next to leading corrections,
mostly those coming from the extended subprocess
kinematics. The method of $k_T$ factorization gives
a reasonable description of the experimental data
up to Tevatron collider energy~\cite{Zotov, Baranov, Sh1}.
The new data on the charm and beauty production
at the LHC energies opens a new opportunity
to compare the theory with the experiment
at the high energies up to $\sqrt s = 13$~TeV.

The heavy quark production at the high energy
goes mainly via gluon fusion
in the small $x$ region.
Here there are no reasons
to neglect the gluon transverse momentum $q_T$
with respect to the relative momentum
of the quark pair $p_T$. At the very high
energies and $p_T\gg m_Q$, $m_Q$ is the quark mass,
the main contribution
to the cross sections comes from
the momenta $q_T \sim p_T$~\cite{Sh2},
which points to the $k_T$ factorization
as a natural tool to deal with it.

In the present paper we give the description of $p_T$
distributions of the charm and beauty mesons produced
at various rapidity intervals,
keeping in mind that the meson distributions
are similar to the heavy quark distributions
\footnote
{In $e^+e^-$ annihilation the charm and beauty
mesons are produced via fragmentation of the incident
heavy quark, so the outcome meson spectra are softer
than the spectra of the heavy quarks.
In the hadron interactions there is a similar way
for the heavy quarks with relatively high $p_T$
to fuse with the light antiquark from
the same shower. However there exists
an alternative possibility to recombine with
the antiquark
originating from another independent branch of
the hadronization.
Thus the produced meson spectra can differ
from the spectrum of fragmentation mesons,
in particular, can be more hard.
This effect is discussed in detail
e.g. in the paper~\cite{Likh}}.

\section{Heavy quark production in $k_T$ factorization}

The cross sections of hard processes
in hadron-hadron interactions
is written in $k_T$ factorization
as the convolutions of the squared
matrix elements of the sub-process calculated
within the perturbative QCD with the parton
distributions in the colliding hadrons,
\begin{eqnarray}
\sigma_{pp}
\,&=&\,\frac{1}{64\pi^2} \frac{1}{s^3}
\int\,
d^{\,4}p_1d^{\,4}p_2\,\delta(x_1y_1s+p_{1T}^2-m_Q^2)
\delta(x_2y_2s+p_{2T}^2-m_Q^2)
\nonumber \\
\label{spp}
&\times &\,d^{\,2} q_{1T}d^{\,2} q_{2T}
\delta^{\,(4)} (q_{1T} + q_{2T} - p_{1T} - p_{2T})\\
&\times &\,\frac{\alpha_s(q^2_1)}{q_1^4}
\frac{\alpha_s (q^2_2)}{q_2^4} f_g(y,q_{1T},\mu)
f_g(x,q_{2T},\mu)\,
\vert\, T(g^*+g^*\,\to\, Q+\overline Q)\,\vert^2.
\nonumber
\end{eqnarray}
The quarks momenta are decomposed here
in Sudakov manner along the momenta
of the protons, $p_A$ and $p_B$, $p_A^2=p_B^2 \simeq 0$,
$2p_A\cdot p_B=s$,
and the transverse momenta $p_{1,2T}$:
\begin{equation}
\label{p12}
p_{1,2} = x_{1,2} p_B + y_{1,2} p_A + p_{1,2T},
~~~d^{\,4}p_{1,2}\,=\,\frac s2 dx_{1,2} dy_{1,2}
d^{\,2}p_{1,2T}.
\end{equation}
The matrix element corresponds to the lowest
order QCD gluon fusion amplitude
$g^*\,+\,g^*\,\to Q\,+\,\overline Q$
taken for the off shell gluons,
$g^*$, whose virtuality
is due to their transverse momentum,
\begin{eqnarray}
\label{q1q2}
q_1\,=x_1^g p_B + y_1^g p_A+ q_{1T}\,
\simeq \,yp_A + q_{1T},
&& q_2 =x_2^g p_B + y_2^g p_A+ q_{2T} \simeq
\,xp_B + q_{2T},\\
q_1^2\,\simeq \, q_{1T}^2,~~~
q_2^2 \simeq \, q_{2T}^2,\,&&
x\,=\,x_1\,+\,x_2,~~~
y\,=\,y_1\,+\,y_2, \nonumber
\end{eqnarray}
$\alpha_S(q^2)$ is one loop running coupling
constant,
$\alpha_s(q^2) =
4\pi /(b \ln{q^2/\Lambda^2})$, $b=11-2/3n_f$,
$\Lambda\simeq 0.25$~GeV.
Like in the parton model the gluons are mainly aligned
in the directions of the colliding hadrons,
the light cone components
$x_1^g\ll y_1^g$, $y_2^g \ll x_2^g$ are neglected
in the amplitude,
but the transverse momenta are no more negligible
and play a central role in the $k_T$ factorization formalism.

Though the incoming partons become virtual for
non vanishing transverse momenta
the hard scattering is gauge invariant
at least at small $x$ where
$k_T$ factorization works.
If we take the gluon propagator
$D^{\mu\nu}(q)=d^{\mu\nu}(q)/q^2$ in the planar gauge,
$d^{\mu\nu}(q)=\delta^{\mu\nu}+(q^\mu n^\nu + q^\nu n^\mu)/q\cdot n$,
the resulting amplitude turns out to be independent
on the gauge fixing vector $n^\mu$~\cite{kT3}.
The main contribution for the large invariant
energy $\sqrt s$ comes from $\delta^{\mu\nu}$
tensor, or more exactly, from its longitudinal part,
so that
$d^{\mu\nu}(q)\simeq 2/s(p_A^\mu p_B^\nu + p_A^\nu p_B^\mu)$.
This form underlies the factorized expression
(\ref{spp}), in which the incoming gluons have to be taken
as purely longitudinal,
$$
T(g^*+g^*\,\to\, Q+\overline Q)\,=
\,\varepsilon^\mu(q_1)\varepsilon^\nu(q_2)\,
T_{\mu\nu}(g^*+g^*\,\to\, Q+\overline Q),
~~~\varepsilon(q_1)=p_A^\mu,~~ \varepsilon(q_2)=p_B^\mu.
$$

Regarded as a part
of the whole scattering the sub process amplitude
is transverse,
\begin{equation}
\label{trans}
(yp_A\,+\,q_{1T})^\mu
T_{\mu\nu}(g^*+g^*\,\to\, Q+\overline Q)\,=\,0,
\end{equation}
and similarly for the second gluon. Due to this
fact the longitudinal polarizations can be equivalently
replaced with the transverse ones \cite{kT3, Sh2},
$\varepsilon^\mu(q_1)=-q_{1T}^\mu/y$,
$\varepsilon^\nu(q_2)=-q_{2T}^\nu/x$,
\begin{equation}
\label{qT}
T(g^*+g^*\,\to\, Q+\overline Q)\,=
\,\frac{q_{1T}^\mu}{y}\frac{q_{2T}^\nu}{x}\,
T_{\mu\nu}(g^*+g^*\,\to\, Q+\overline Q),
\end{equation}
but it allows for more general polarizations
vectors,
\begin{equation}
\label{Apols}
\varepsilon_A^\mu\,=\,-\frac 1y
(q_{1T} - 2y p_A)^\mu, ~~
\varepsilon_B^\nu\,=\,-\frac 1x
(q_{2T} - 2x p_B)^\nu.
\end{equation}
All these forms are equivalent within
$k_T$ factorization accuracy owing to condition
(\ref{trans}).

If the restrictions $|q_T|\ll |p_T|$ were valid for
the all gluons' and quarks' momenta the matrix element
would turn into the standard parton model expression
for the real gluons, whereas
the $q_{1,2T}$ integrals (\ref{qT}) in the cross section
(\ref{spp}) recover averaging over their helicities.
For $|q_T|\simeq |p_T|\gg m_Q^2$ the matrix element becomes
much more complicated. It looks rather bulky
and quite different from the parton model one.

There is a way, however, to modify the matrix element in a manner
that drastically simplifies it making it closer to the standard
parton model expression. To do it we note at first that $Q\overline
Q$ pair can be produced in the kinematics (\ref{q1q2}) only if
\begin{equation}
\label{sxy}
s_{xy}\,=\,xys \,-\, |(q_{1T}+q_{2T})|^2 \gtrsim 4(m_Q^2+|p_T|^2)
\end{equation}
Assuming that in the integral (\ref{spp})
$x,y \ll 1$ and $x\sim y$
it is natural to conclude that for heavy quarks
$y \gg |q_{1T}|^2/(ys)$, $x \gg |q_{2T}|^2/(xs)$
provided $s$ is large enough.
It allows to substitute
in the relations (\ref{q1q2}) the light cone
components with the values
$x_1^g=-q_{1T}^2/(ys)$ and $y_2^g=-q_{2T}^2/(xs)$.
These components still can be neglected in the matrix
element thereby not affecting the $k_T$ factorization validity.
On the other hand they put the incoming gluons momenta
on the mass shell. Moreover,
the vectors (\ref{Apols}) turn out to be the proper polarizations,
$q_1\cdot \varepsilon_A=0$, $q_2\cdot \varepsilon_B=0$.

Thus having modified incoming momenta one arrives
at the amplitude for the scattering of quasireal particles
in the spirit of Weizsacker-Williams method.
Further, the amplitude of the real
process remains unchanged after
adding to the gluon polarization any vector proportional
to its momentum.
Using this freedom the two polarizations $\varepsilon_{A},
\varepsilon_{B}$ can be replaced with the equivalent ones
$e_{A}/y,e_{B}/x$, such that
$q_1\cdot e_{A,B}=q_2\cdot e_{A,B}=0$.
It brings the matrix
element to the final form,
\begin{equation}
\label{me}
\bigl|T(g^*+g^*\,\to\, Q+\overline Q)\bigr|^2\,=\,
\frac 2{x^2y^2s_p^2}\,
\frac{4 - 9\,z(1-z)}
{3\,(1 - z)^2z^2\,}
\end{equation}
$$
\times\,\biggl\{4\,\bigl[e_A\cdot e_B\,z(1-z)\,s_p
+ 2\,e_A\cdot p_\perp\,e_B\cdot p_\perp\bigr]^2
+ (1 - z)z\,s_p^2\,e_A\cdot e_A\,e_B\cdot e_B\biggr\}.
$$
This is just the squared parton model
Born amplitude $g+g\,\to\, Q+\overline Q$
except for it is not averaged over $e_A$, $e_B$
polarizations.
Here $s_p = 2\,q_1\cdot q_2$ is the pair invariant energy,
$p_\perp$ is the quark relative momentum,
$p_\perp\cdot q_1 = p_\perp\cdot q_2 = 0$.
The constraints  $q_1+q_2=p_1+p_2$, $p_1^2=p_2^2=m_Q^2$
are resolved for $q_{1,2}^2=0$ as
$$
p_1\,=\,z\,q_1\,+\,(1-z)\,q_2\,+\,p_\perp,~~~~
p_2\,=\,(1-z)\,q_1\,+\,z\,q_2\,-\,p_\perp,
$$
$$
d^{\,4}p_1 d^{\,4}p_2\,\delta^{(4)}(q_1+q_2-p_1-p_2)
\delta(p_1^{\,2}-m_Q^2)\,\delta(p_2^{\,2}-m_Q^2)
$$
$$
=\,\frac 12 dp_\perp^{\,2} dz\,
\delta\bigl[|p_\perp|^{\,2}-z(1-z)s_p-m_Q^2\bigr],
$$
leaving $p_\perp$ as a single independent variable.
The radial $|p_\perp|$
integral is convenient to do in terms of the variable $z$,
ranging in the interval
$(1-\rho)/2\le z \le (1+\rho)/2$,
$\rho^2 = 1 -4m_Q^2/s_p$.

It is worth to point out that $p_\perp$
is not orthogonal to the protons' momenta
$p_A$ and $p_B$.
The conventional parton model kinematics
looks like it is
rotated with respect to the direction of the colliding
protons in their center of mass frame.

\begin{figure}[htb]
\centering
\includegraphics[width=.3\hsize]{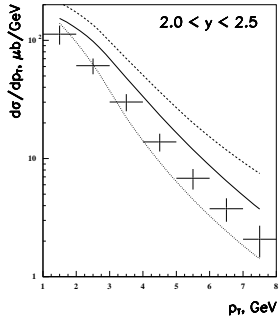}
\vskip -1.0 cm
\caption{\footnotesize
The cross section
of $c$ quark production at $\sqrt s = 7$~TeV
in the rapidity interval
$2\le y \le 2.5$ calculated with 3 values
of the factorization scale: $\mu^2=m_c^2$,
$m_c$ is the $c$-quark mass,
(dotted line), $\mu^2=m_T^2$ (dashed line)
and $\mu^2=m_T^2/2$ (solid line),
$m_T$ is the $c$-quark transverse mass.
The experimental points are taken
from ~\cite{Aaij:2013mga}.
}
\label{mu}
\end{figure}

To make the "external" integrals over $x$, $y$,
$q_{1,2T}$ the explicit expressions are needed:
$$
e_A\cdot e_A\,=\,q_{1T}^2\,=\,-|q_{1T}|^2,~~~
e_B\cdot e_B\,=\,q_{2T}^2\,=\,-|q_{2T}|^2,~~~
e_A\cdot e_B = |q_{1T}||q_{2T}|\frac{[a\, b\,]}{s_p},
$$
$$
[a\,b\,]\,=\,\frac 1{s_{xy}}\biggl[(q_{1T}^2q_{2T}^2 + s_{xy}^2)
\frac{q_{1T}\cdot q_{2T}}{|q_{1T}||q_{2T}|}
- 2|q_{1T}||q_{2T}|s_{xy}\biggr],
$$
$$
s_p\,=\,\frac 1{s_{xy}}\bigl[q_{1T}^2q_{2T}^2 + s_{xy}^2
+ 2q_{1T}\cdot q_{2T}\, s_{xy}\bigr],
$$
where $s_{xy}$ is defined in (\ref{sxy}).
The cross section of the heavy quark pair production
now reads
\begin{eqnarray}
\sigma\,&=&\,\frac 1{16\pi s}\int\frac{dq_1}{q_1^4}
\alpha_S(q_1^2)\frac{dq_2}{q_2^4}
\alpha_S(q_2^2)\frac{dx}{x^2}\frac{dy}{y^2}\,
dz\, d\phi\, d\theta  \\
&&\times\,\bigl|\,T(g^*+g^*\,\to\, Q+\overline Q)\bigr|^2
\,f_g(y,q_1,\mu)f_g(x,q_2,\mu).\nonumber
\end{eqnarray}
In this expression $q_{1,2}=|q_{1,2T}|$,
the angle $\phi$ is defined as
$q_{1T}\cdot q_{2T}=-|q_{1T}||q_{2T}|\cos\phi$,
the variable $z$ and the angle $\theta$ represent
the integral over $p_\perp$.

The integrals over small $q_{1,2T}$ reproduce
averaging of the matrix element over gluon helicities
occurring in the standard parton model. For the not small
momenta they involve averaging over the orientations
of the two dimensional plane where the parton model kinematics
is relevant or, in other words, where the quarks' relative
momentum $p_\perp$ lies. In contrast to the conventional
case it is not the plane orthogonal to the colliding hadrons,
therefore the transverse quark momenta $p_{1,2T}$
in (\ref{p12}) are not directed along $p_\perp$.
To pass to the center of mass frame for
the colliding hadrons it suffices to reexpress
$p_{A}$, $p_B$ vectors through the momenta $q_{1,2}$
and the polarizations $e_A$, $e_B$.

\begin{figure}[htb]
\vskip -0.5 cm
\includegraphics[width=.3\textwidth]{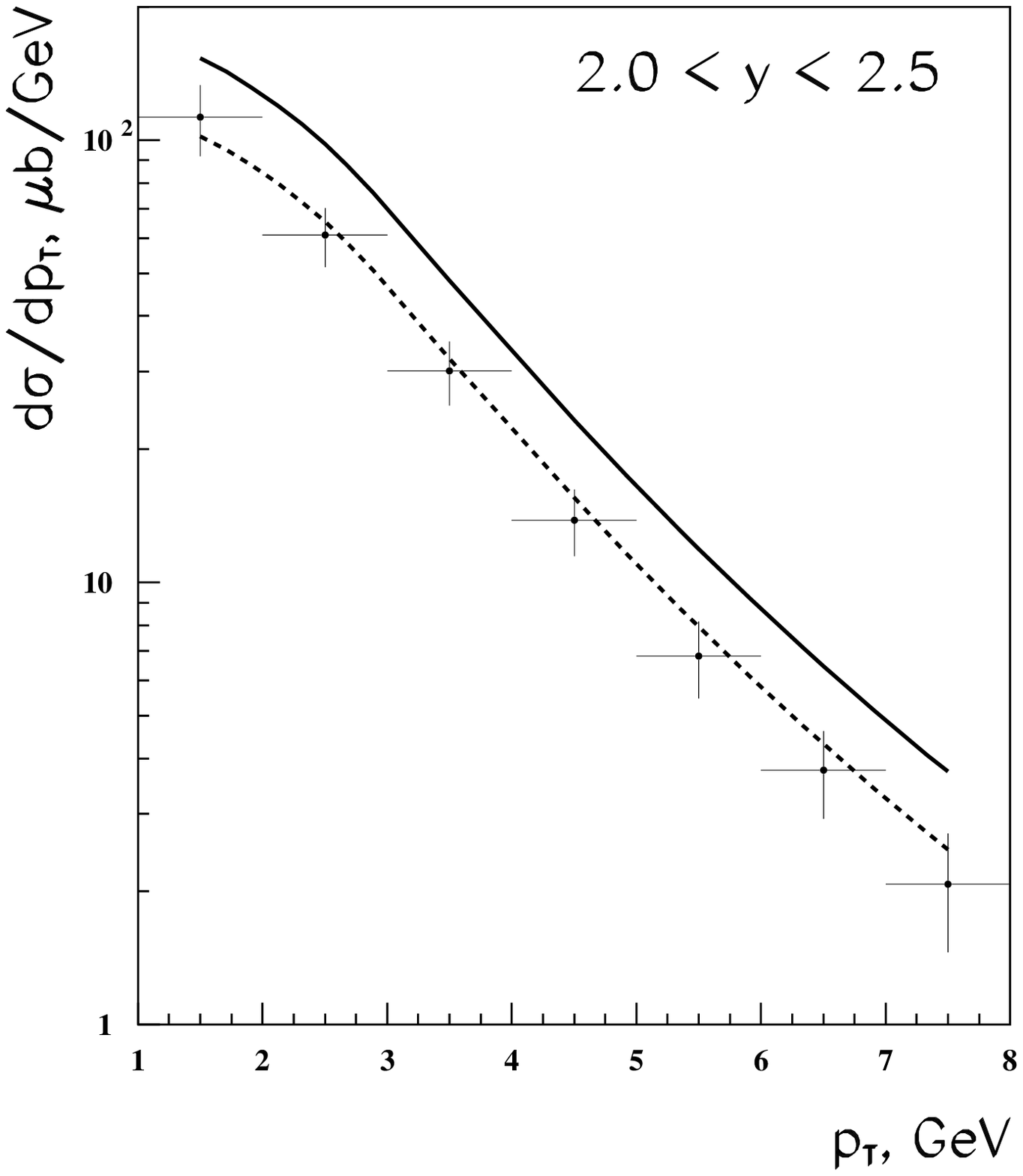}
\includegraphics[width=.3\textwidth]{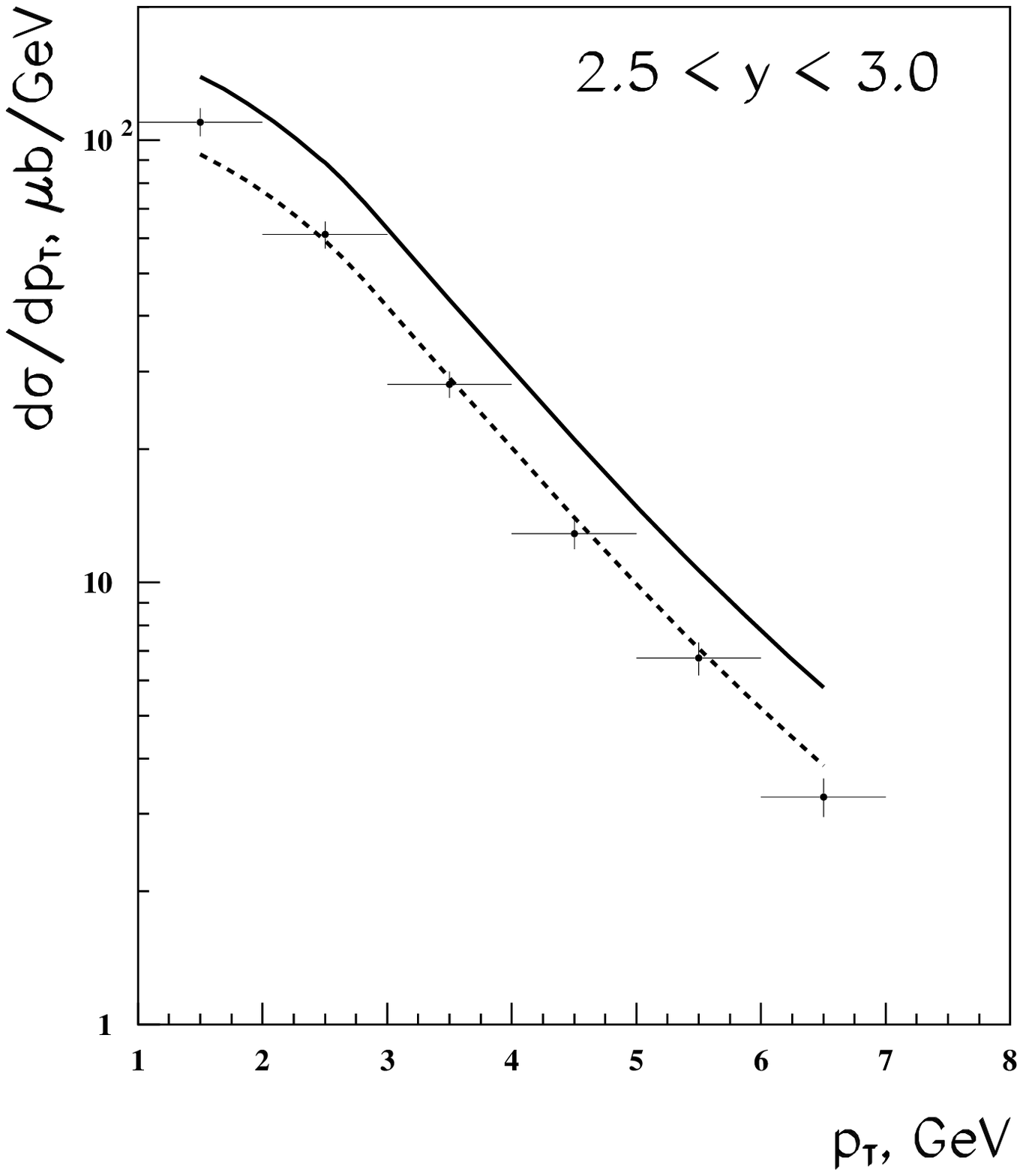}
\includegraphics[width=.3\textwidth]{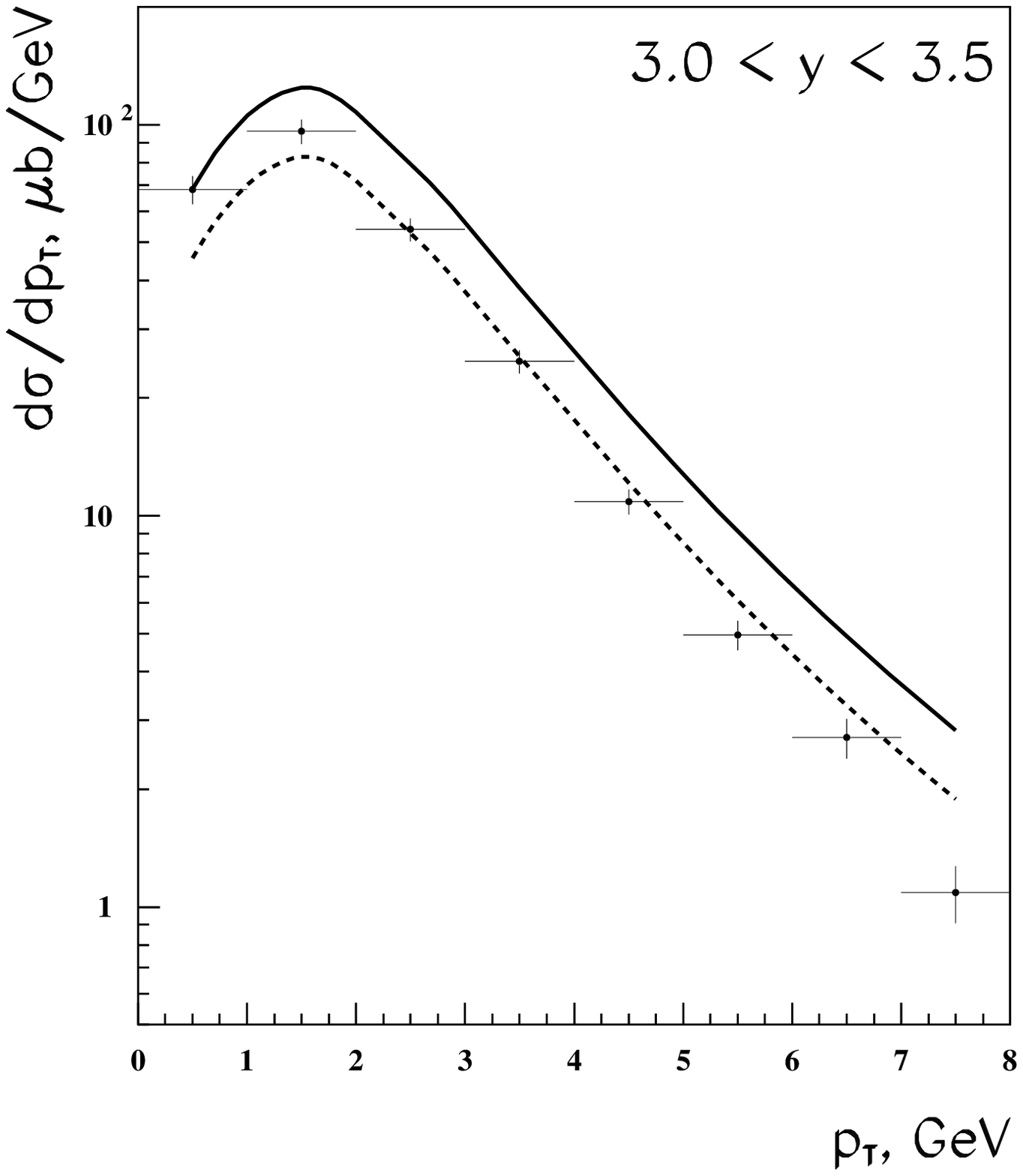}
\vskip -0.4cm
\includegraphics[width=.3\textwidth]{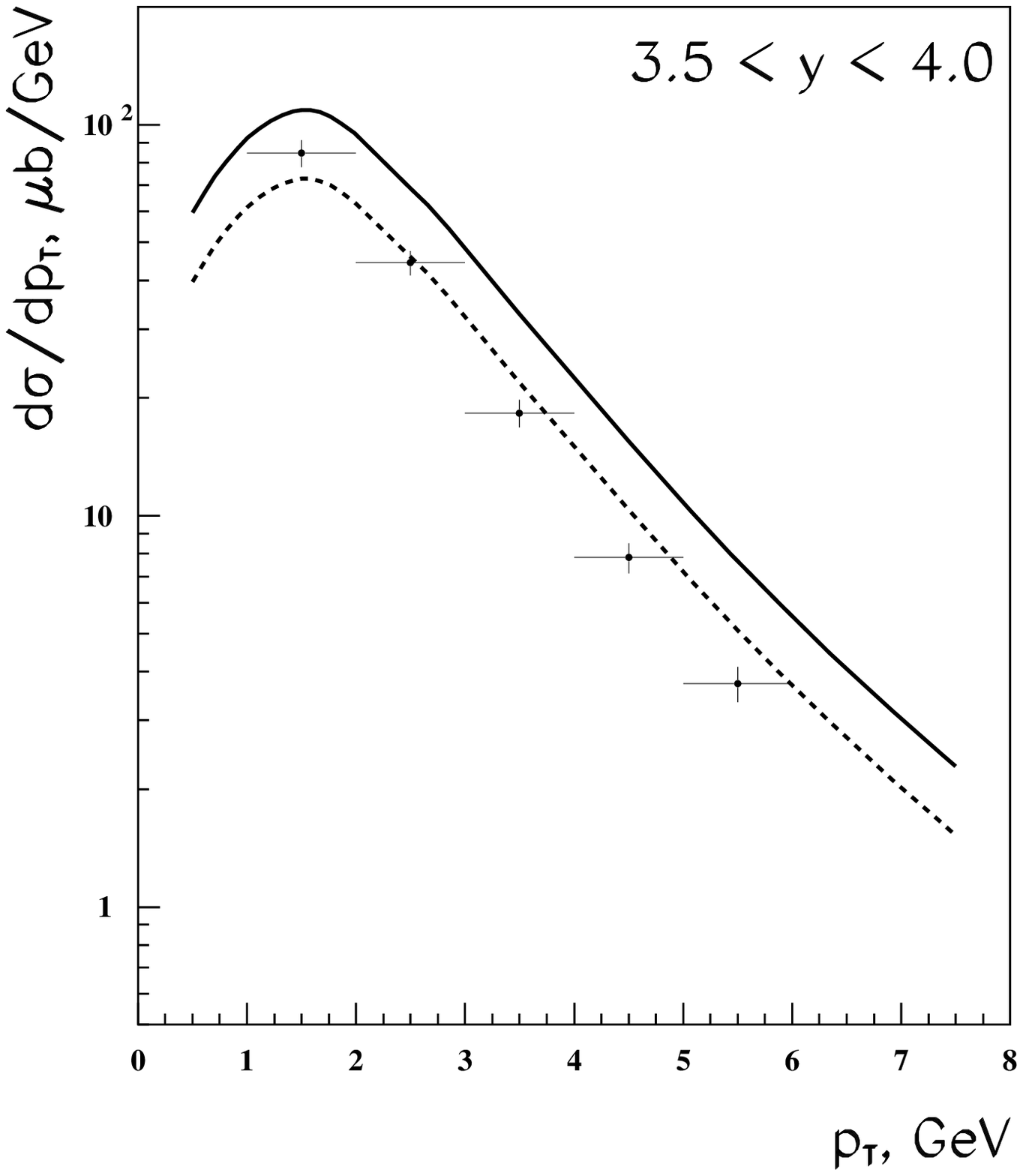}
\includegraphics[width=.3\textwidth]{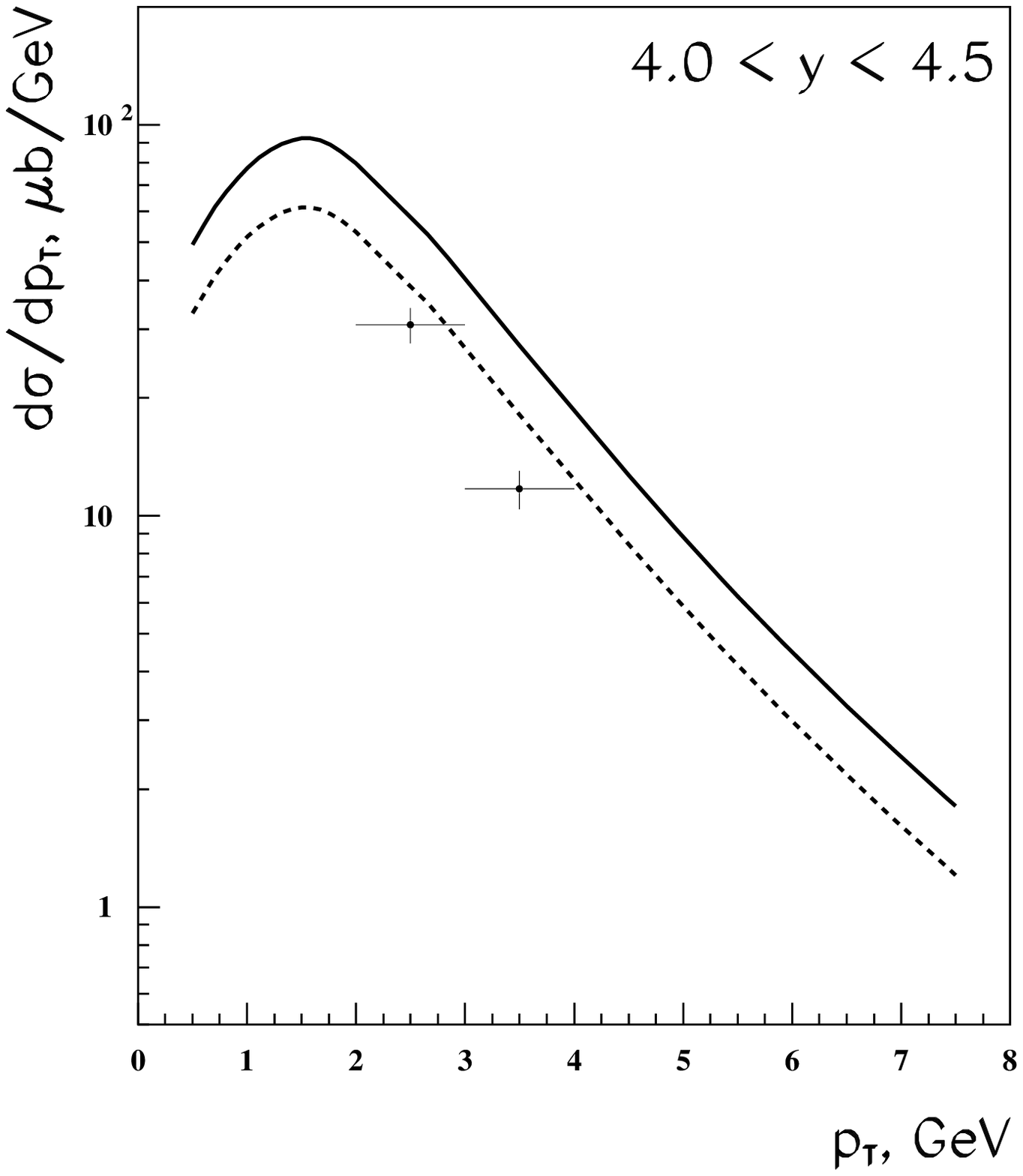}
\vskip -1.5 cm
\caption{\footnotesize
$p_T$ dependence of the charm production
in various rapidity regions
at $\sqrt s=7$~Tev. The solid curves are for
the total charm production cross section,
the dashed curves are for the charm meson production only.
The experimental data are taken from~\cite{Aaij:2013mga}.}
\label{c7}
\end{figure}

The unintegrated parton distribution
$f_g(x,q_T,\mu)$, entering the cross section (\ref{spp}),
determines the probability to find a gluon
initiating the hard process with the longitudinal
momentum fraction $x$ and the transverse momentum $q_T$.
The factorization scale $\mu$ sets an upper momentum bound
for the parton to be included into the distribution function.
The partons carrying larger momenta have to be treated
as participating in a rescattering, that gives rise
to the NLO corrections or to the jet production etc.
To find the function $f_g(x,q_T,\mu)$ on the basis
of the conventional (integrated) gluon density
$G(x,Q^2)$ we employ Kimber-Martin-Ryskin
(KMR) approach \cite{KMR1, KMR2},
\begin{equation}
\label{f_g}
f_g(x,q_T,\mu) =  T_g(q_T,\mu)
\left[\frac{\alpha_S(q_T^2)}{2 \pi}
\int^{\Delta}_x P_{gg} (z)\frac xz
G\left(\frac xz,q_T^2 \right) dz
\right],
\end{equation}
where $P_{gg} (z)$ is the LO DGLAP gluon-gluon
splitting function. It is singular for $z\to 1$,
the singularity coming from the real soft gluon
emission is regulated by the cutoff $\Delta$.
The singularity is canceled
by the virtual loop corrections that are collected
in the survival probability for the gluon to evolve
untouched up to the factorization scale,
\begin{equation}
\label{Sud}
T_g(q_T,\mu) = \exp \left[ -\int^{\mu^2}_{q^2_T}
\frac{\alpha_S(p_T^2)}{2\pi} \frac{dp^2_T}{p^2_T}
\int^{\Delta}_0 P_{gg}(z)dz \right].
\end{equation}
The regulator is taken here as
$\Delta = \mu/\sqrt{\mu^2 + q_T^2}$,
the numerical results do not rather
sensitive to its particular form~\cite{Sh1}.
Since the main contribution comes to the integral
(\ref{f_g}) from $z\sim 1$ we put in the integrand
$x/\!z\, g(x/z,q_T^2)\approx x g(x/z,q_T^2)$ to avoid
too singular behavior at $z\sim 0$.

The structure functions are unknown in the infrared
domain of the small momenta $q_{1,2T}^2$. That is why
the contributions from $|q_{1,2T}^2|<Q_0^2$ and
$|q_{1,2T}^2|>Q_0^2$, $Q_0^2\sim 1$~GeV$^2$, are
calculated separately. When $|q_{T}^2|<Q_0^2$
the unintegrated distribution is replaced with
the usual structure function,
$f_g(x,q_T,\mu)=x G(x,Q_0^2)T(Q_0,\mu)$,
multiplied by the survival probability
$T(Q_0,\mu)$
not to have transverse momenta larger than $Q_0^2$
(see \cite{Sh1}).
The parametrization \cite{GRV} having a rather simple
analytical form is taken
for the gluon structure function $G(x,Q^2)$.

\section{Comparison with the experimental data}

We start from the data on the charm production obtained
at the LHC at the energy $\sqrt s = 7$~TeV
because they contain the cross sections
of the production of charm $D^0$, $D^+$, $D_s^+$ mesons
together with $\Lambda_c^+$ baryon. It makes it more
suitable to compare with since the produced $c$ quark
can fragment into the mesons as well into the baryons.

There are two basic parameters in the calculation --
the mass of the heavy quark and the factorization scale.
We take the same $c$ and $b$ masses as in our previous
paper~\cite{Sh1}:
$$
m_c\,=\, 1.4~{\rm GeV},~~~ m_b\,=\, 4.6~{\rm GeV}.
$$

The factorization scale $\mu^2$ separates the partons
participating in the process from those responsible
for the evolution of the structure function.
It should be taken to be of the order
of the typical hardness of the reaction.
The role of $\mu^2$ becomes more important at high energies
in the small $x$ region, where the structure function grows.

The cross section $d\sigma/dp_T$
of $c$ quark production calculated
at $\sqrt s = 7$~TeV in the rapidity interval
$2\le y \le 2.5$ are presented in Fig.~\ref{mu} for
three different values of the factorization scale: $\mu^2=m_c^2$,
$\mu^2=m_T^2$ and $\mu^2=m_T^2/2$, where $p_T$ is the quark
transverse momenta, $m_T^2=m_c^2+|p_T|^2$. The curve for
$\mu^2=m_T^2$ is seen to decrease with $p_T$ too slow
while for $\mu^2=m_c^2$ it decreases too fast
so almost everywhere it lies below the data
also shown in the Fig.~\ref{mu}.
The curve with
$\mu^2=m_T^2/2$ better fits the data.
It has roughly the same slope though
goes uniformly above the experimental points.
It is the value $\mu^2=m_T^2/2$ that has been taken
for the further calculations.

\begin{figure}[htb]
\centering
\includegraphics[width=.3\hsize]{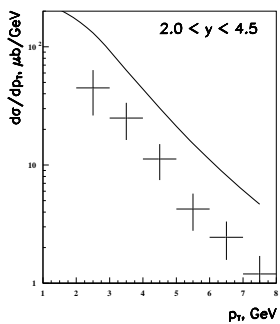}
\vskip -1 cm
\caption{\footnotesize
The cross section
of $\Lambda_c$ baryon production at $\sqrt s = 7$~TeV
in the rapidity interval $2\le y \le 4.5$.
The experimental data are taken from~\cite{Aaij:2013mga}.}
\label{Lmc}
\end{figure}
The factorization scale significantly controls
in what extent the parton rescattering affects
the calculation output for a given scheme.
Thus the choice of $\mu^2$ not only influences
the perturbative NLO corrections
but may help to achieve the better description
of the hadronization stage.
\begin{figure}[htb]
\vskip -0.5 cm
\includegraphics[width=.3\textwidth]{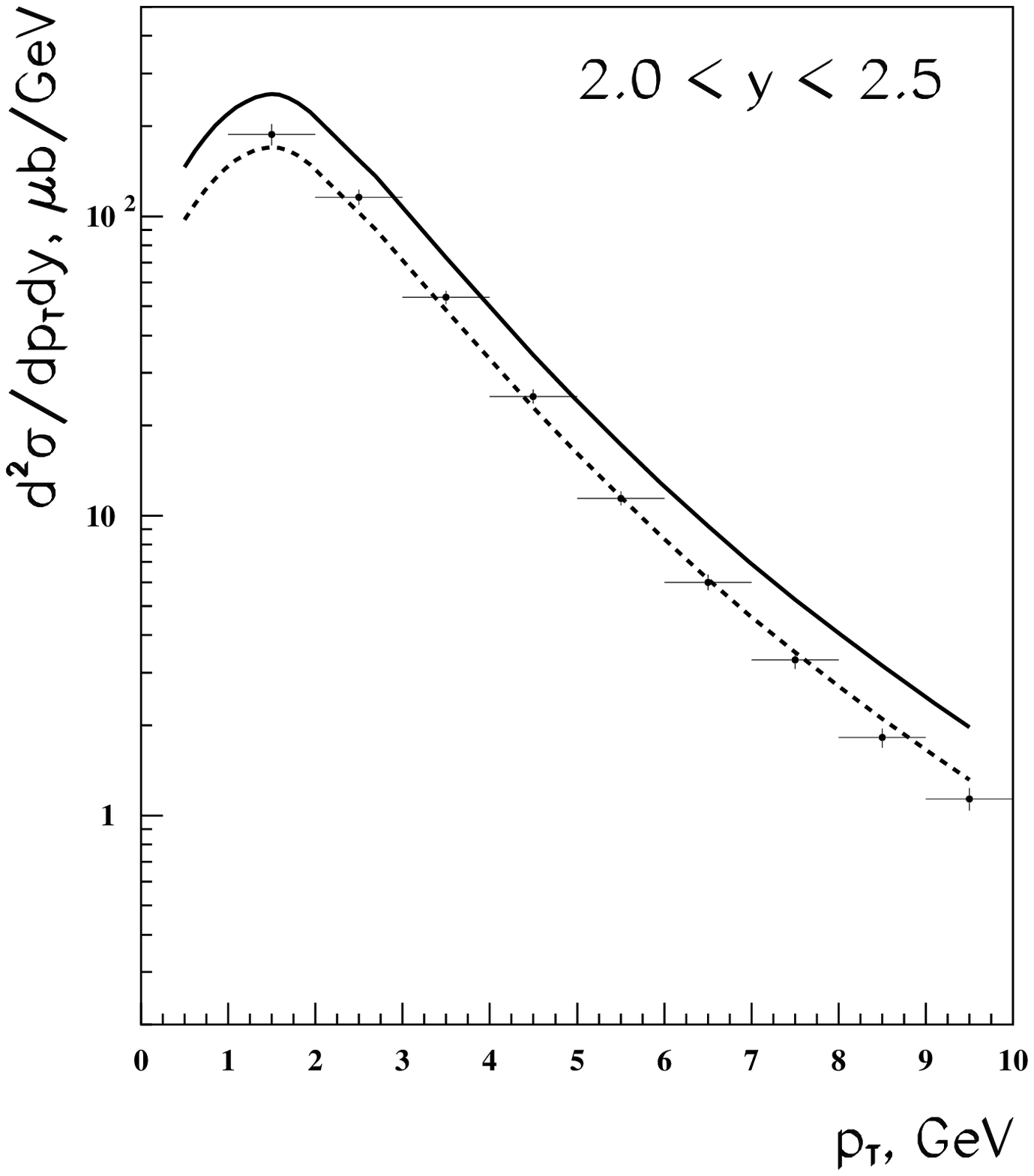}
\includegraphics[width=.3\textwidth]{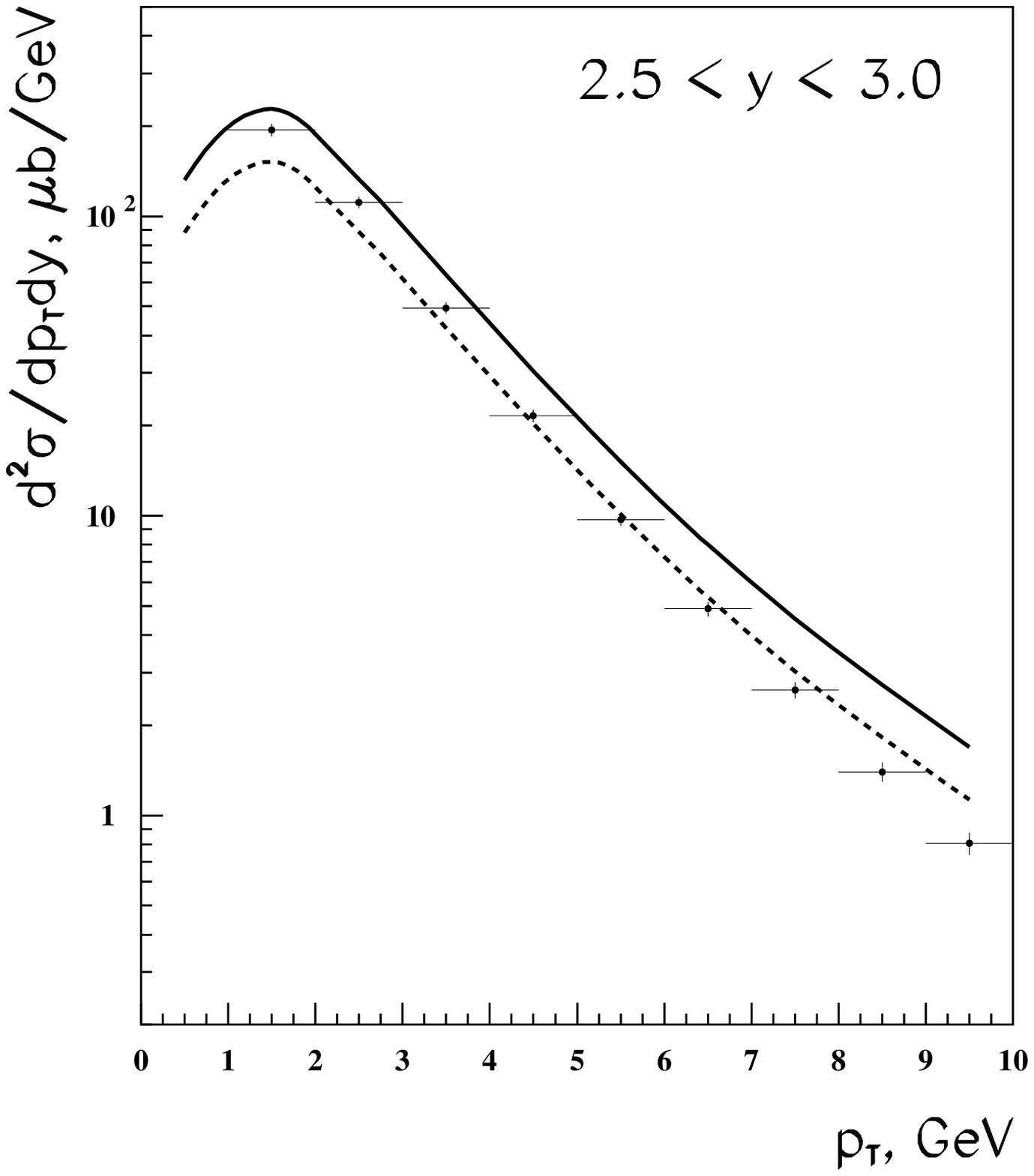}
\includegraphics[width=.3\textwidth]{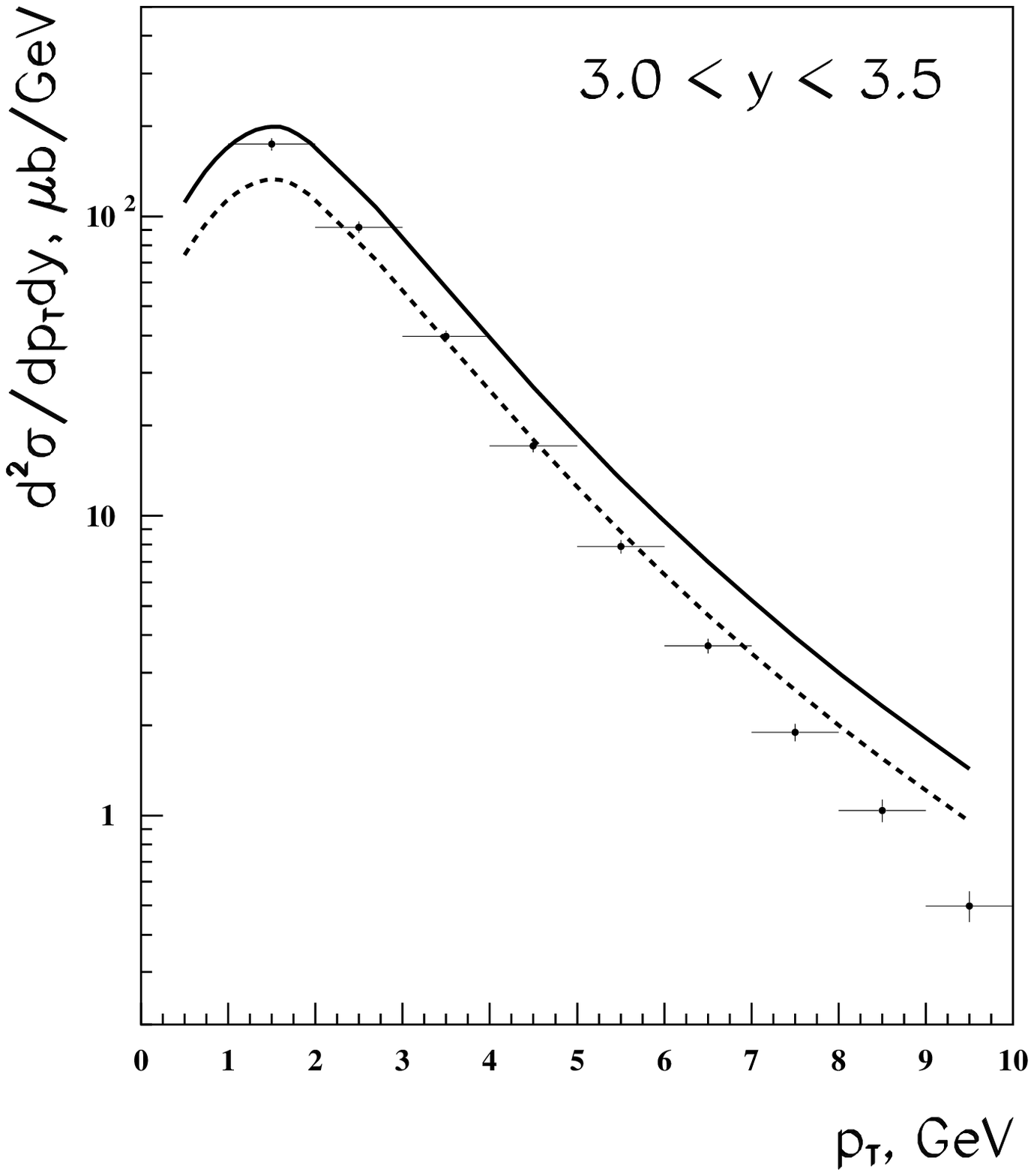}
\vskip -0.4cm
\includegraphics[width=.3\textwidth]{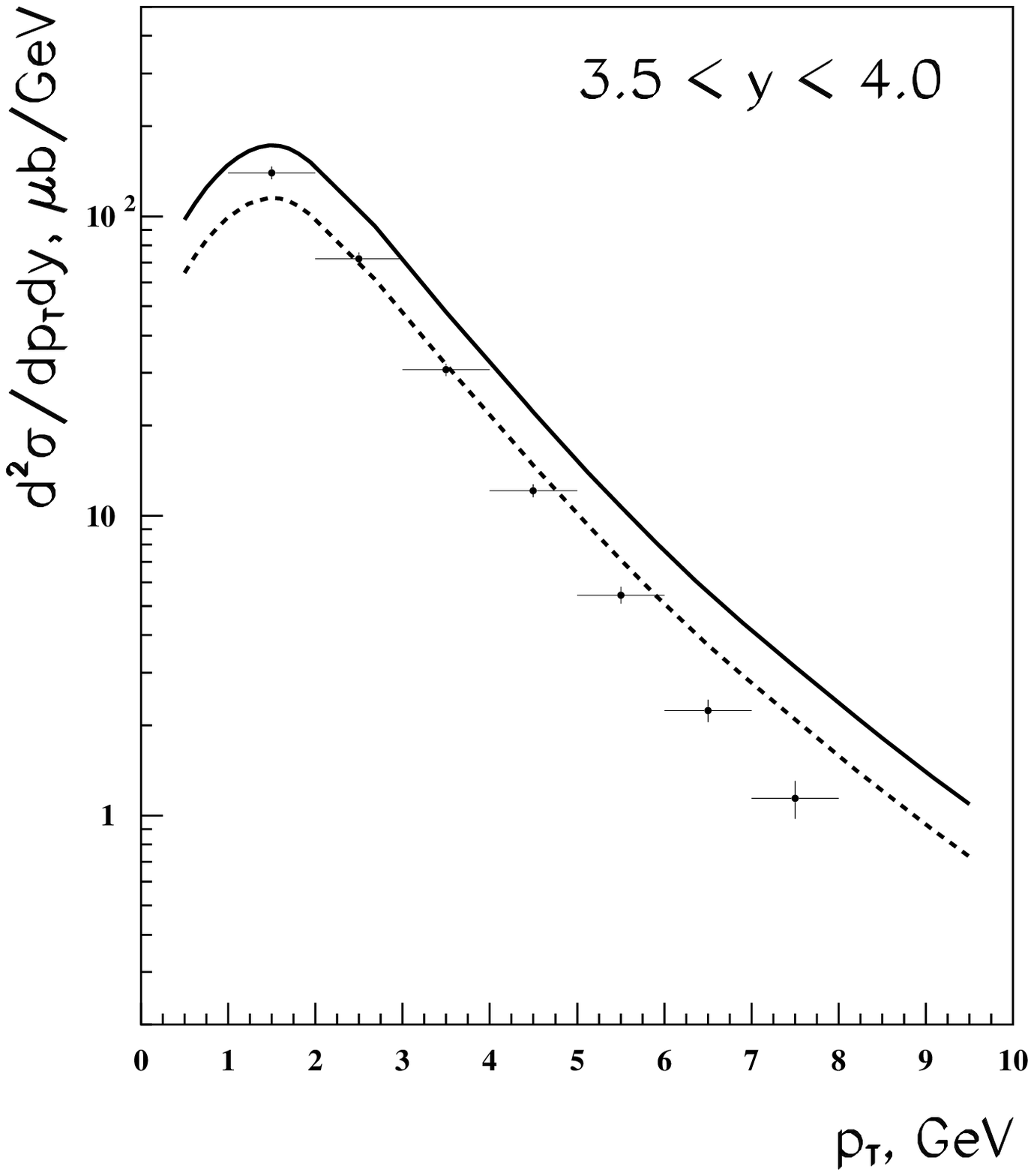}
\includegraphics[width=.3\textwidth]{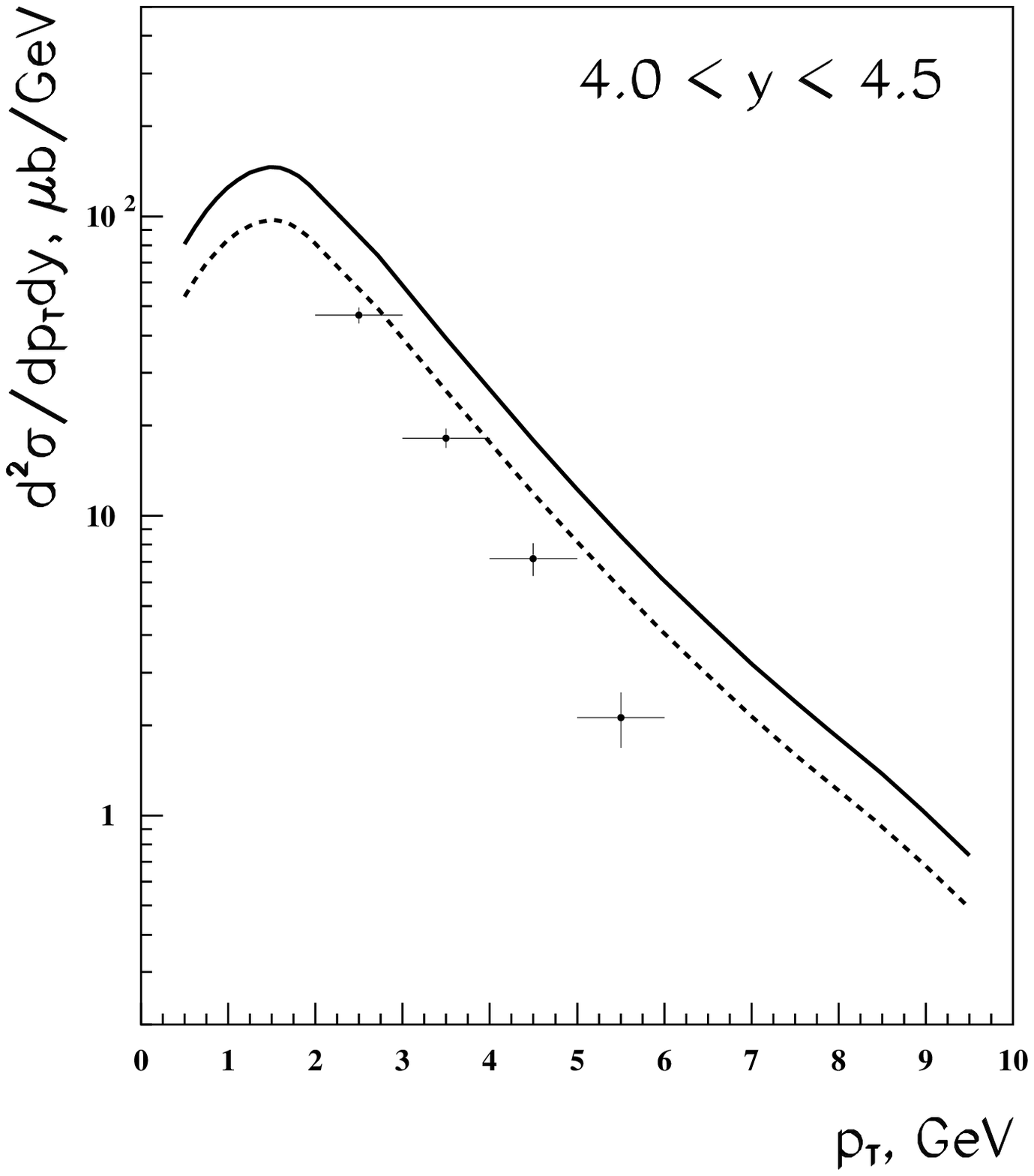}
\vskip -1.5 cm
\caption{\footnotesize
$p_T$ dependence of the charm production
in various rapidity regions
at $\sqrt s=5$~Tev. The solid curves are for
the total charm production cross section,
the dashed curves are for the charm meson production only.
The experimental data are taken from~\cite{Aaij:2016jht}.}
\label{c5}
\end{figure}

The normalization exceeding of the calculated results above
the experiment can be explained by the fact that only
the charm mesons $D^0$, $D^+$, $D_s^+$ are mostly detected
while the $c$ quark production implies the subsequent
fragmentation also into the charm baryons such as $\Lambda_c$ etc.
To estimate the correct normalization one can employ
the simple quark combinatorics
\cite{Anisovich-Schehter,Anisovich}.
Assuming
\cite{Anisovich-Schehter,Anisovich}.
the same probabilities for the charm quark
to couple with a quark or antiquark one gets after
the first fusion
$$
c\,+\,(1/2\,q +1/2 \,\bar q)\,\to\, 1/2\, c\,q
\,+\, 1/2 M_c.
$$
Then the second fusion gives
$$
1/2\, c\,q\, +\, (1/2\, q+1/2\,\bar q)\,\to \,
 1/4\, B_c + 1/4\,c +  1/4 \,M,
$$
where $M_c$, $B_c$ are the charmed meson and baryon,
$M$ is the sea meson. Thus
the charmed mesons and baryons should be produced
in the proportion 2:1, that is
\begin{equation}
\label{M_c}
c\,\to \, 2/3\, M_c \,+\, 1/3\, B_c.
\end{equation}
As was shown in Ref~\cite{AVK}
this relation is in reasonable agreement with
the experimental data obtained in $e^+e^-$ annihilation.
However the multiplicity of secondary baryons
is significantly smaller than 1/3 for the pion nucleon
collisions~\cite{Sh3}.
Nevertheless we use the ratio (\ref{M_c})
as an upper boundary for baryon production
for the absence of another theoretical
models.

In all the following figures we show the results
for $c$ quark production
by the solid curves and the results for the charmed meson
production estimated according to eq.(\ref{M_c})
by the dashed curves.

\begin{figure}[htb]
\vskip -1.cm
\includegraphics[width=.3\textwidth]{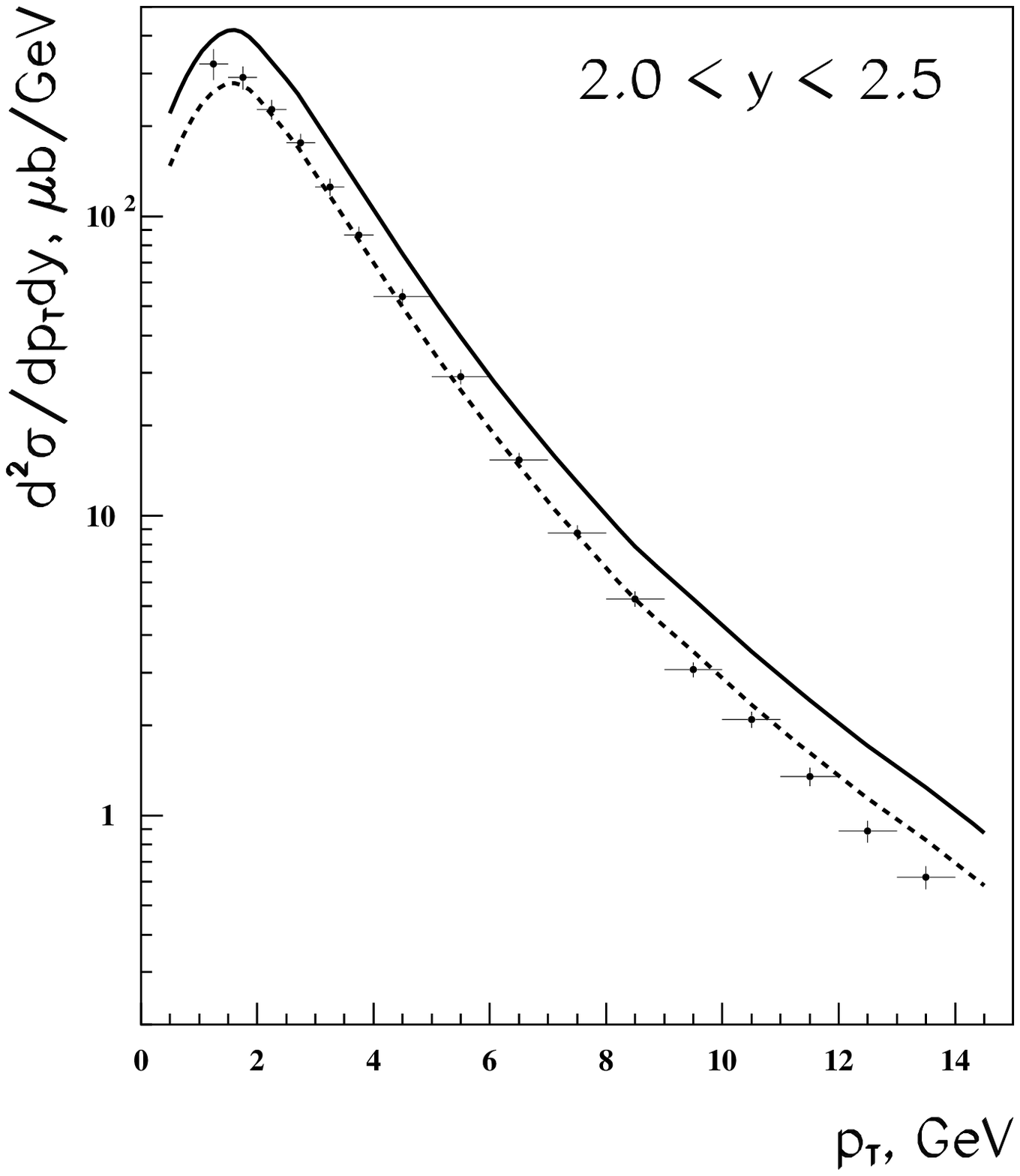}
\includegraphics[width=.3\textwidth]{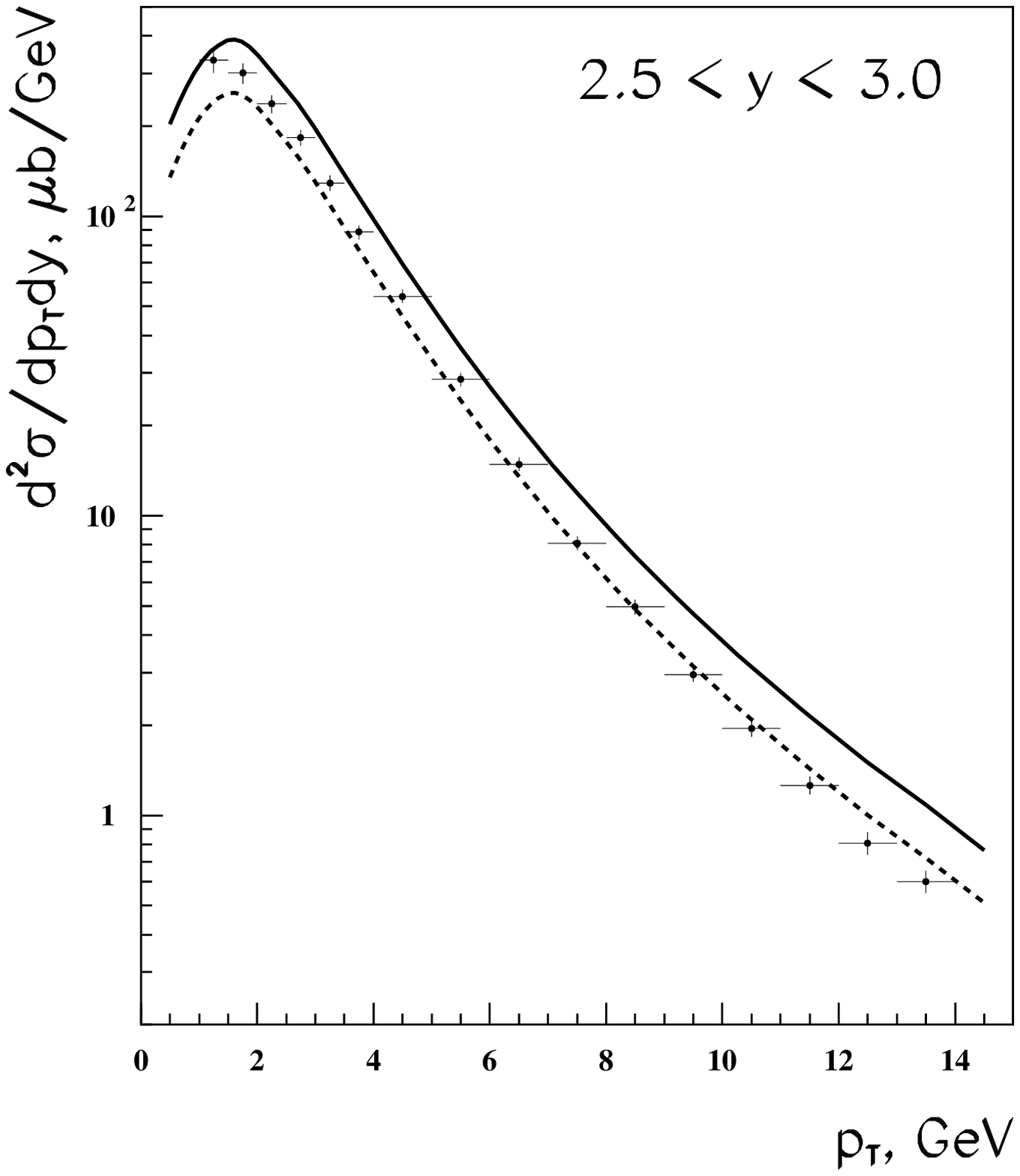}
\includegraphics[width=.3\textwidth]{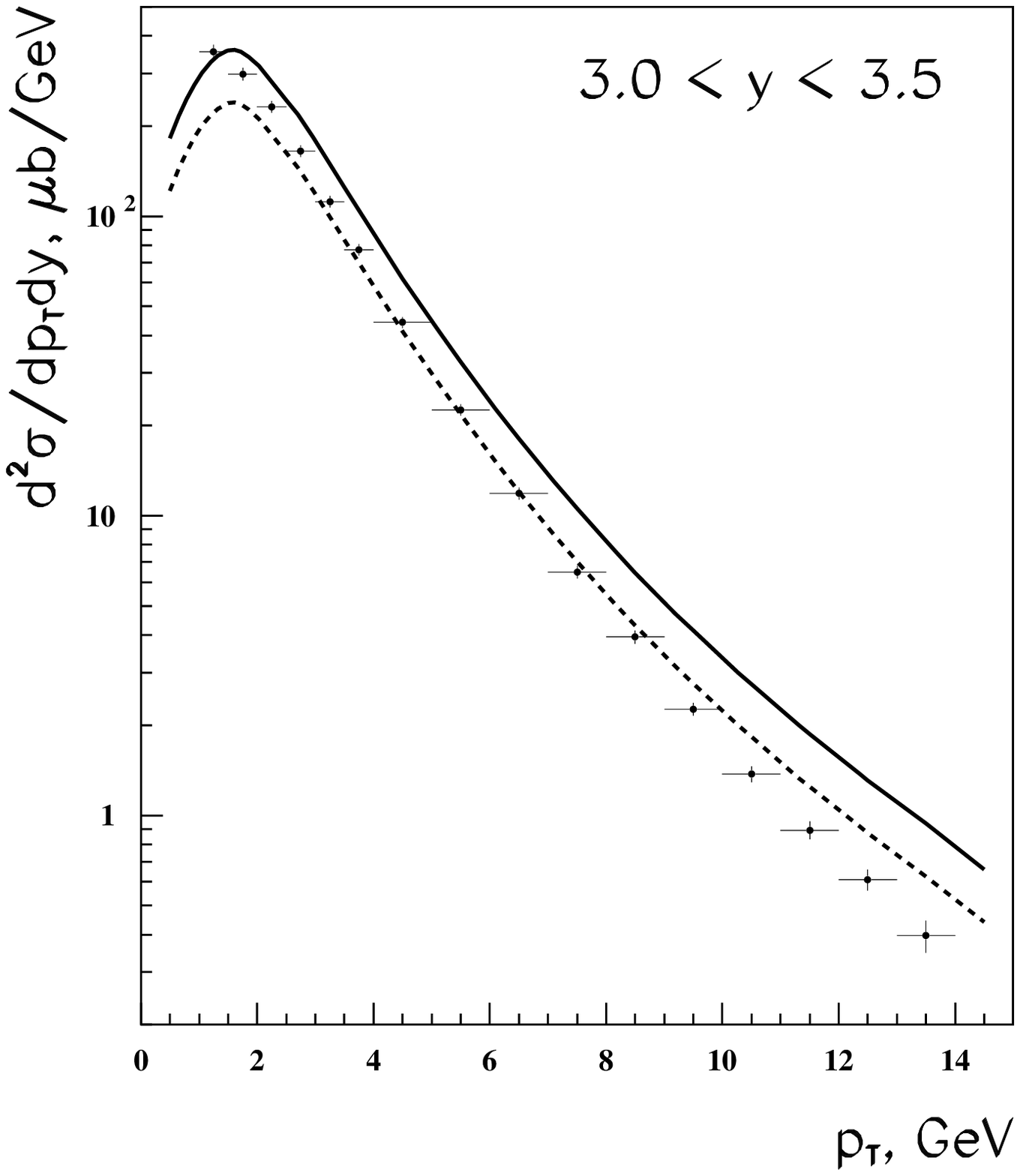}
\vskip -0.4cm
\includegraphics[width=.3\textwidth]{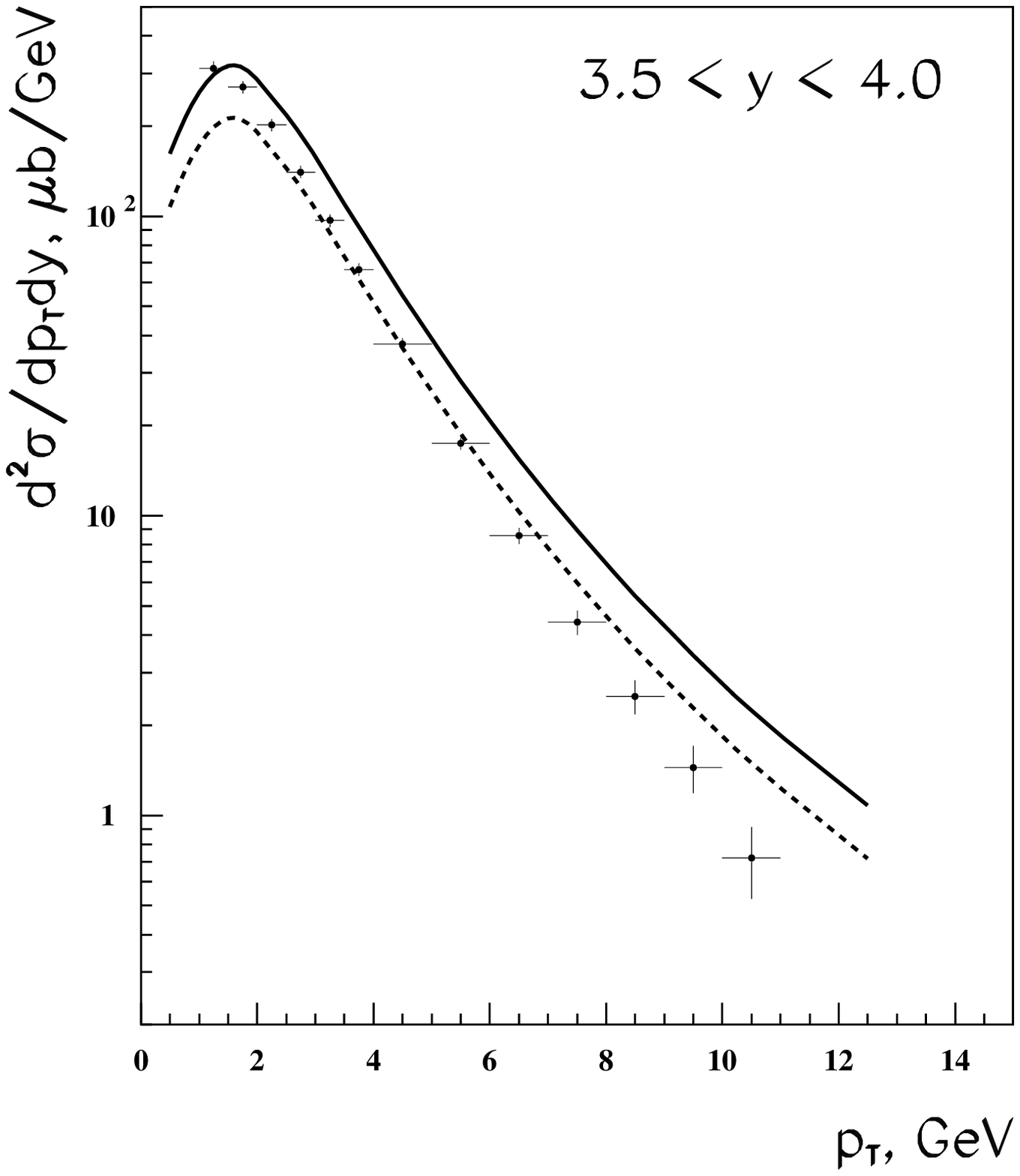}
\includegraphics[width=.3\textwidth]{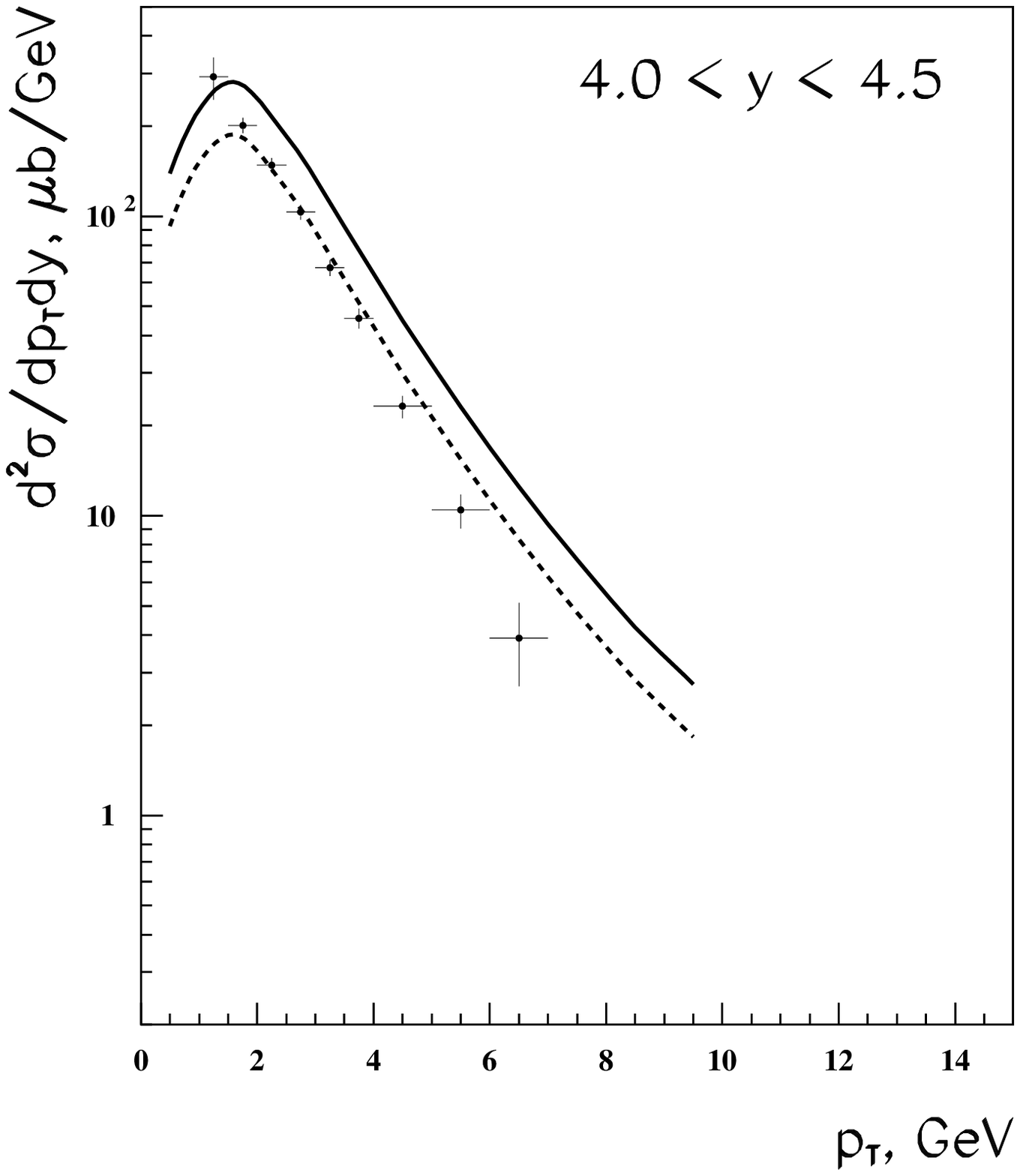}
\vskip -1.5 cm
\caption{\footnotesize
$p_T$ dependence of the charm production
in various rapidity regions
at $\sqrt s=13$~Tev. The solid curves are for
the total charm production cross section,
the dashed curves are for the charm meson production only.
The experimental data are taken from~\cite{Aaij:2015bpa}.}
\label{c13}
\end{figure}

The comparison with the experimental data
on $p_T$ dependence of charm production at
$\sqrt s=7$~TeV in various
rapidity intervals are presented in Fig.~\ref{c7}.
The solid curves are higher than the experimental
points but the dashed curves for the charm meson
production are in reasonable agreement in all cases.
Some discrepancy at the large rapidities can be explained
by overestimated small $x$ region in the GRV structure function
\cite{GRV}.

\begin{figure}[htb]
\centering
\includegraphics[width=.3\textwidth]{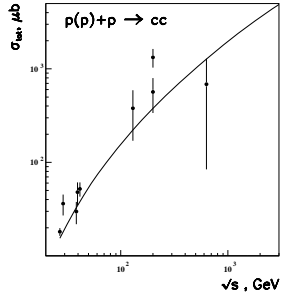}
\vskip -0.5 cm
\caption{\footnotesize
Total cross section of charm production
in $pp$ and $\bar p p$ collisions. The experimental
points are taken from
\cite{Lourenco, Abreu, Botner, Adcox:2002cg,
Adare:2006hc, Bielcik:2006ef}.}
\label{ctot}
\end{figure}

\begin{figure}[htb]
\vskip -1.cm
\includegraphics[width=.3\textwidth]{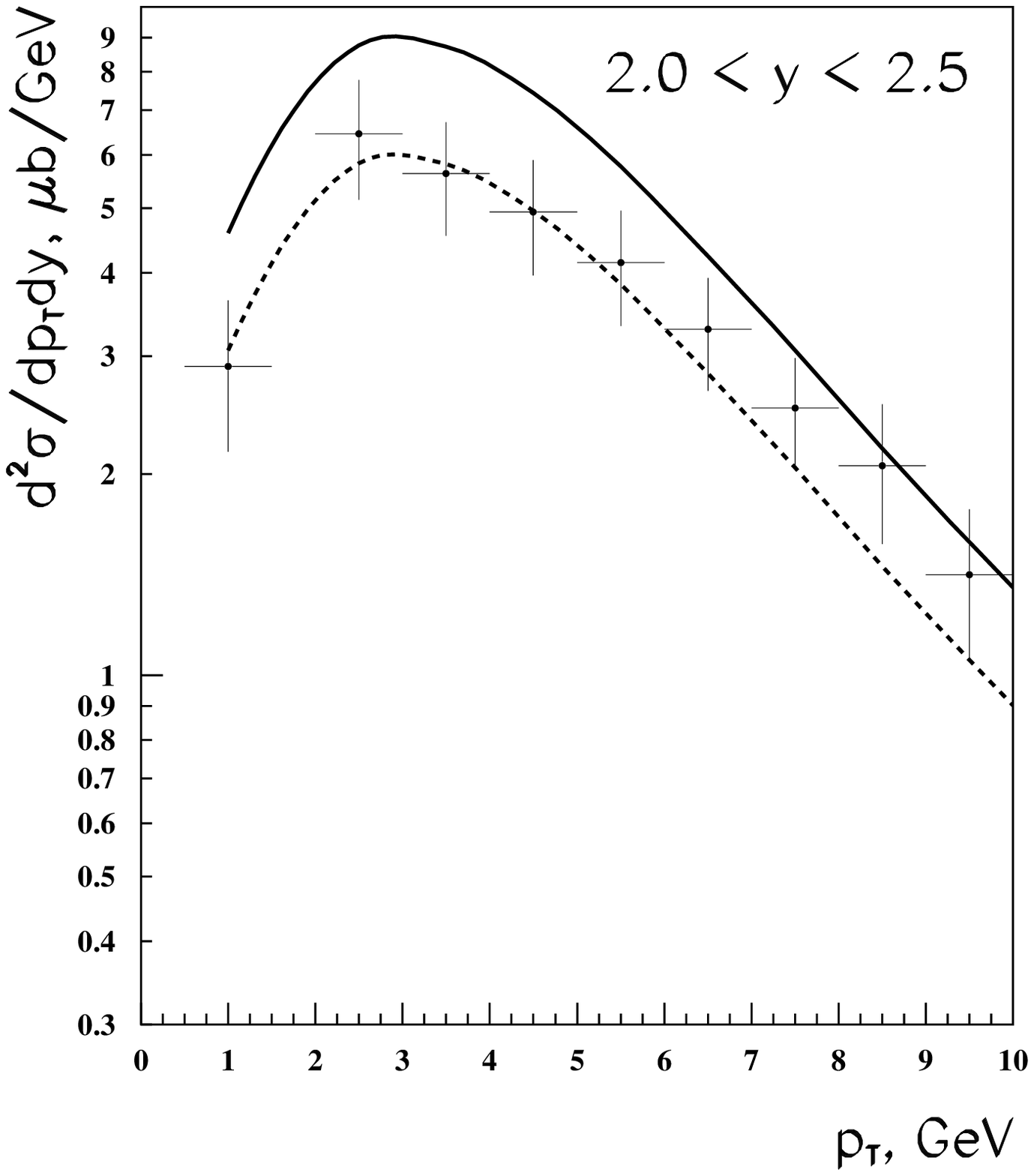}
\includegraphics[width=.3\textwidth]{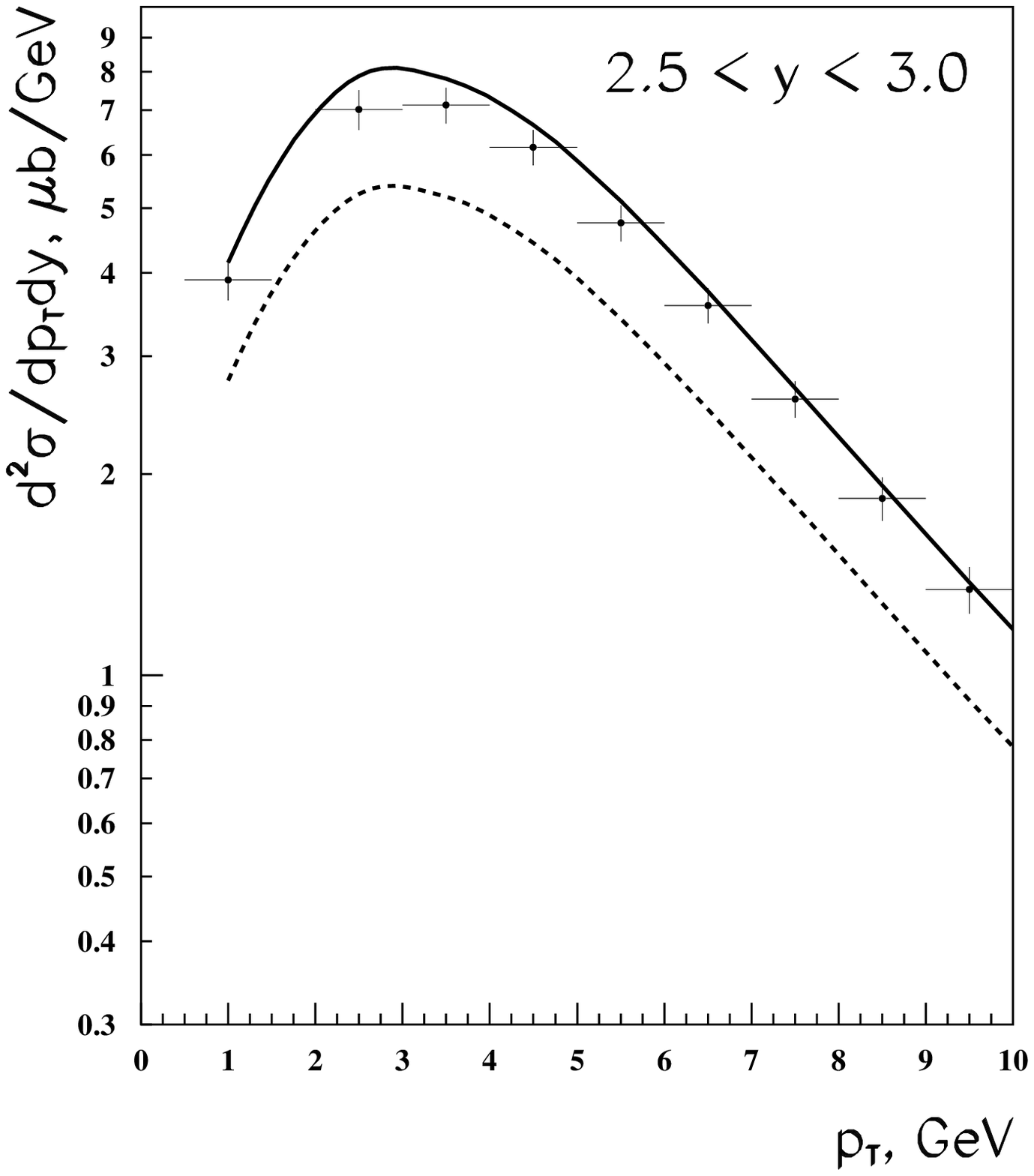}
\includegraphics[width=.3\textwidth]{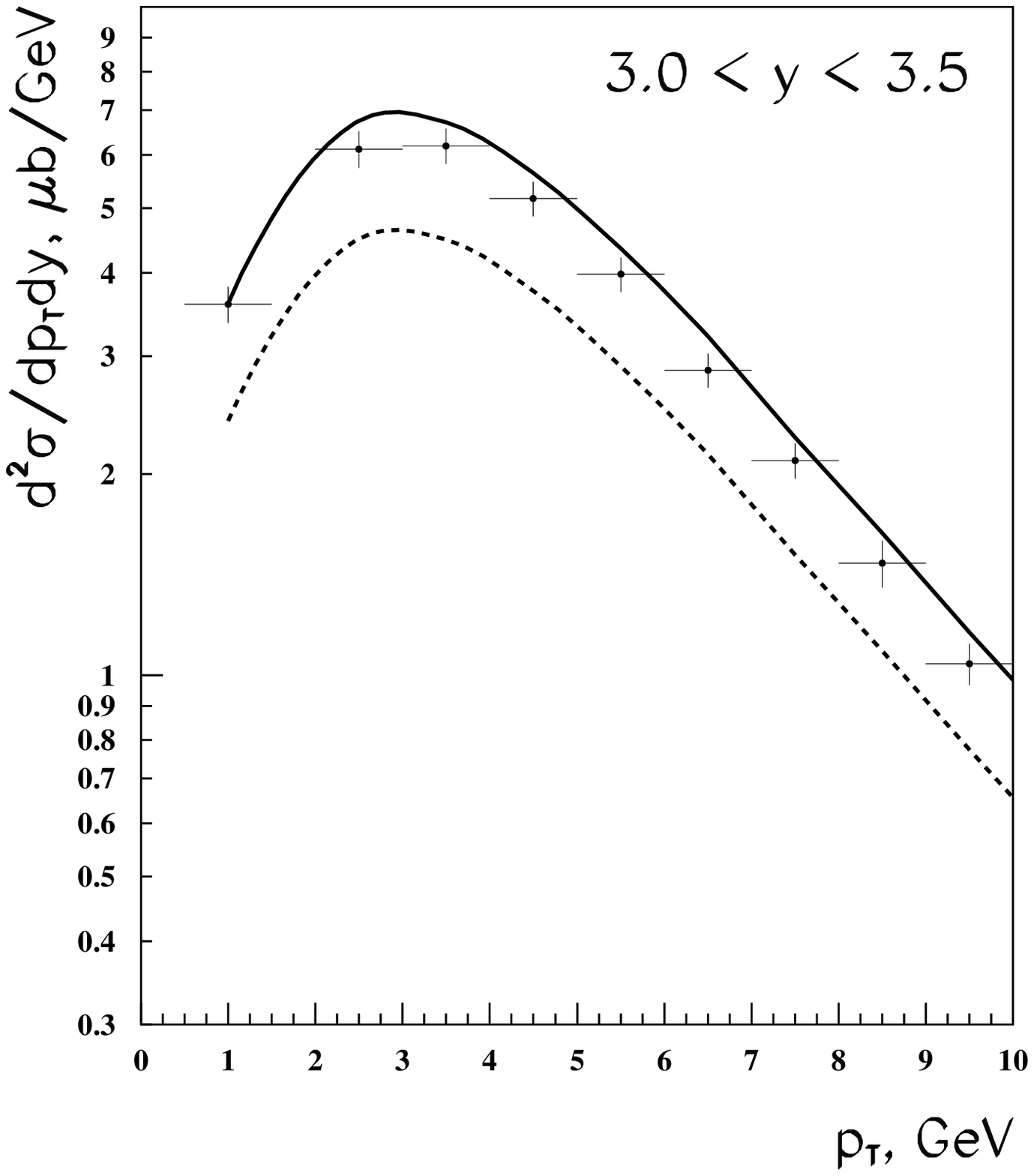}
\vskip -0.4cm
\includegraphics[width=.3\textwidth]{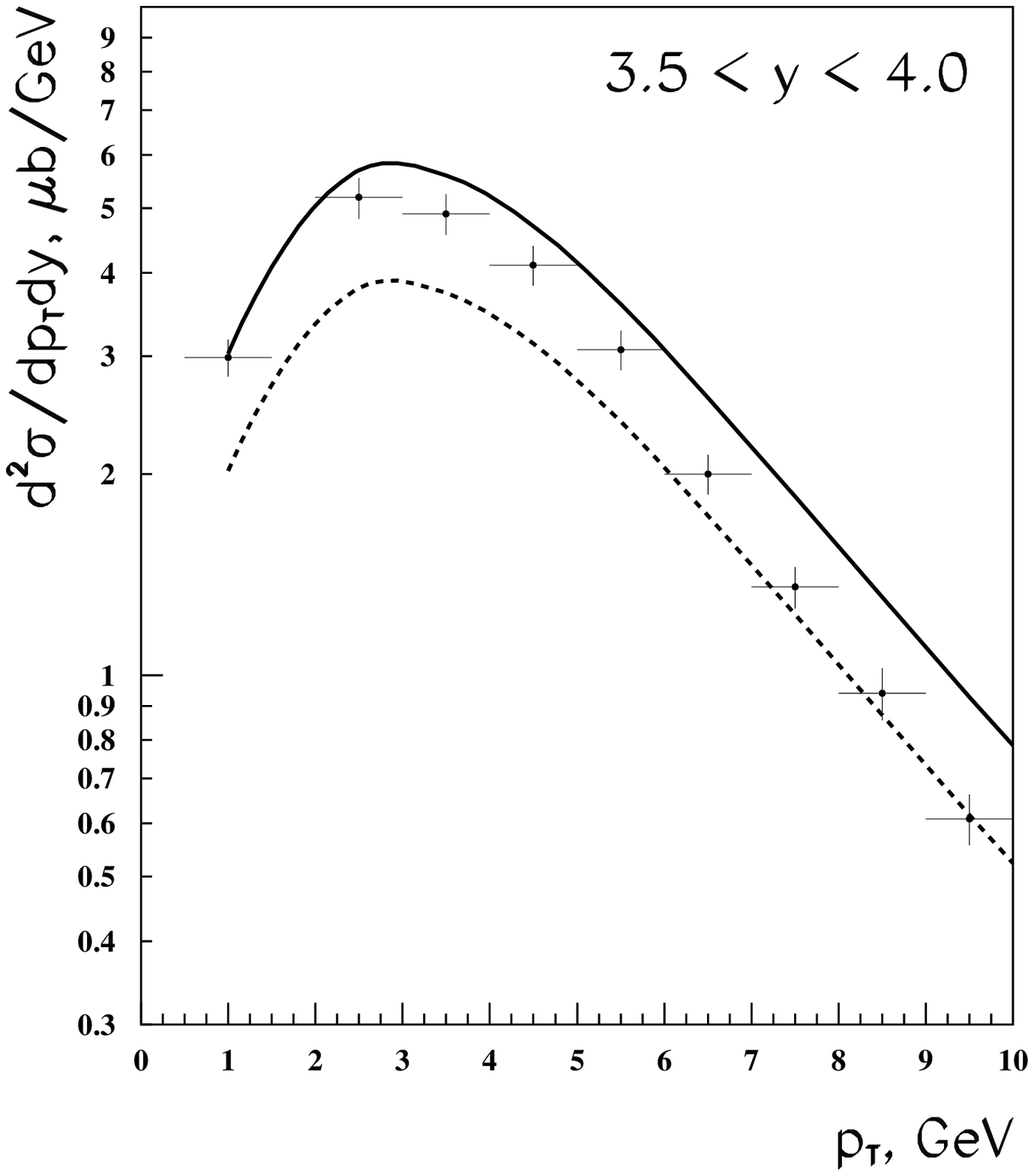}
\includegraphics[width=.3\textwidth]{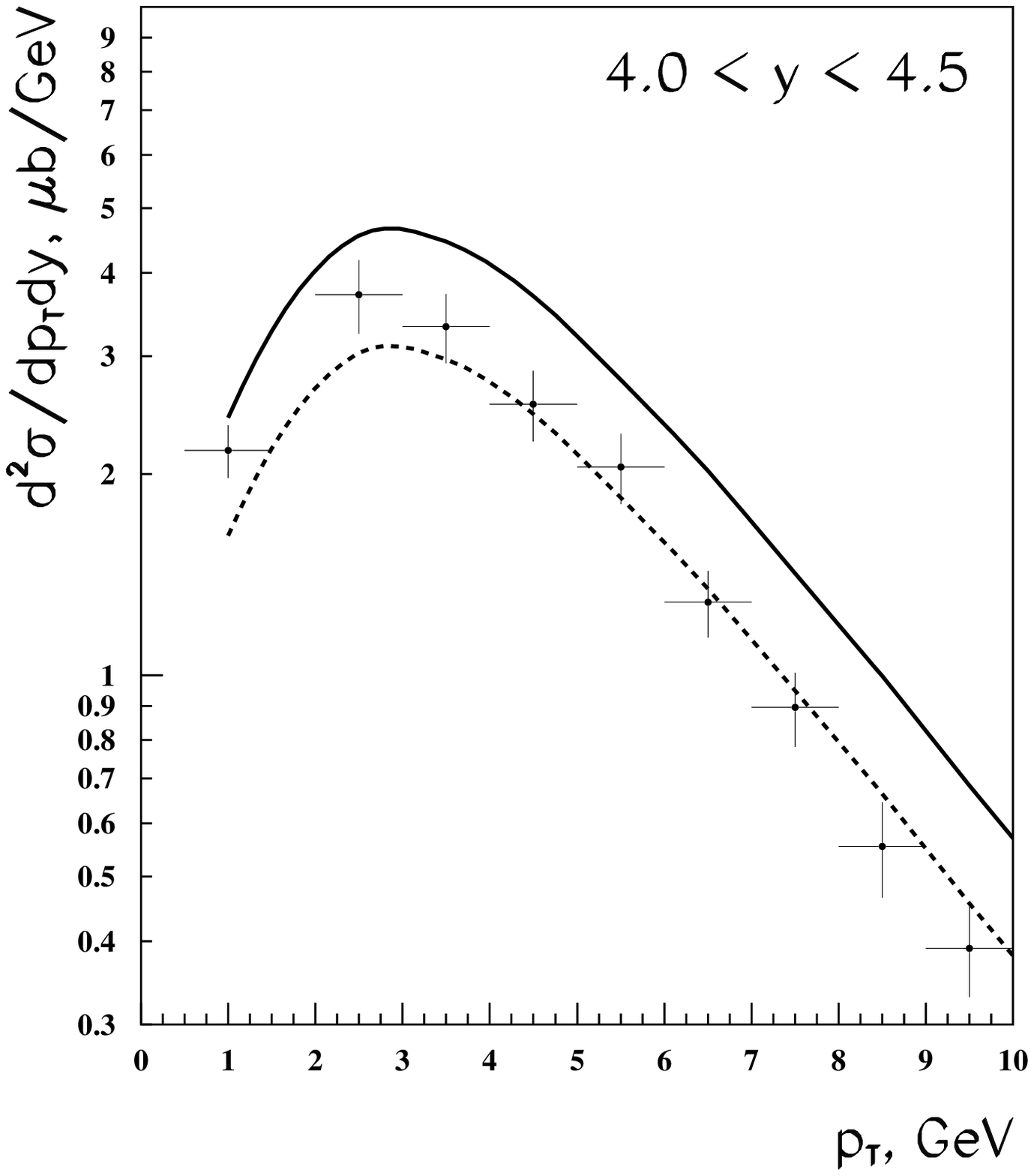}
\vskip -1.5 cm
\caption{\footnotesize
$p_T$ dependence of the beauty production
in various rapidity regions
at $\sqrt s=7$~Tev. The solid curves are for
the total beauty production cross section,
the dashed curves are for the beauty meson production only.
The experimental data are taken from~\cite{Aaij:2013noa}.}
\label{b7}
\end{figure}

To check up the ratio of the charmed meson/baryon outcome
the experimental data on $\Lambda_c$ production at 7~TeV
are presented in Fig.~\ref{Lmc} together with our results
for the charm baryon production estimated as
1/3 of the total one. The evaluated curve
is higher than the experimental points.
A possible reasons could be in another undetected
charm hadrons or in a violation of the quark combinatorics.

There also exist the experimental data for the charm mesons
production at 5~TeV and 13~TeV as the functions of $p_T$
in various rapidity regions. They are compared with our
calculations in Fig.~\ref{c5} and Fig.~\ref{c13}.
Again the dashed curves demonstrate reasonable agreement
with the data at the both energies.

Our calculation scheme yields the reasonable values for the total
charm production cross section at the lower energies. They are shown
in Fig.~\ref{ctot}.

The results of our approach extended within the
same assumptions to the beauty production are presented in
Fig.~\ref{b7}. They show the smaller than 1/3 probability for the
$b$ quark to fragment into baryons. Thus the cross section of $B$
meson production is perhaps closer to the total cross section of $b$
quark production.
\begin{figure}[htb]
\centering
\includegraphics[width=.3\textwidth]{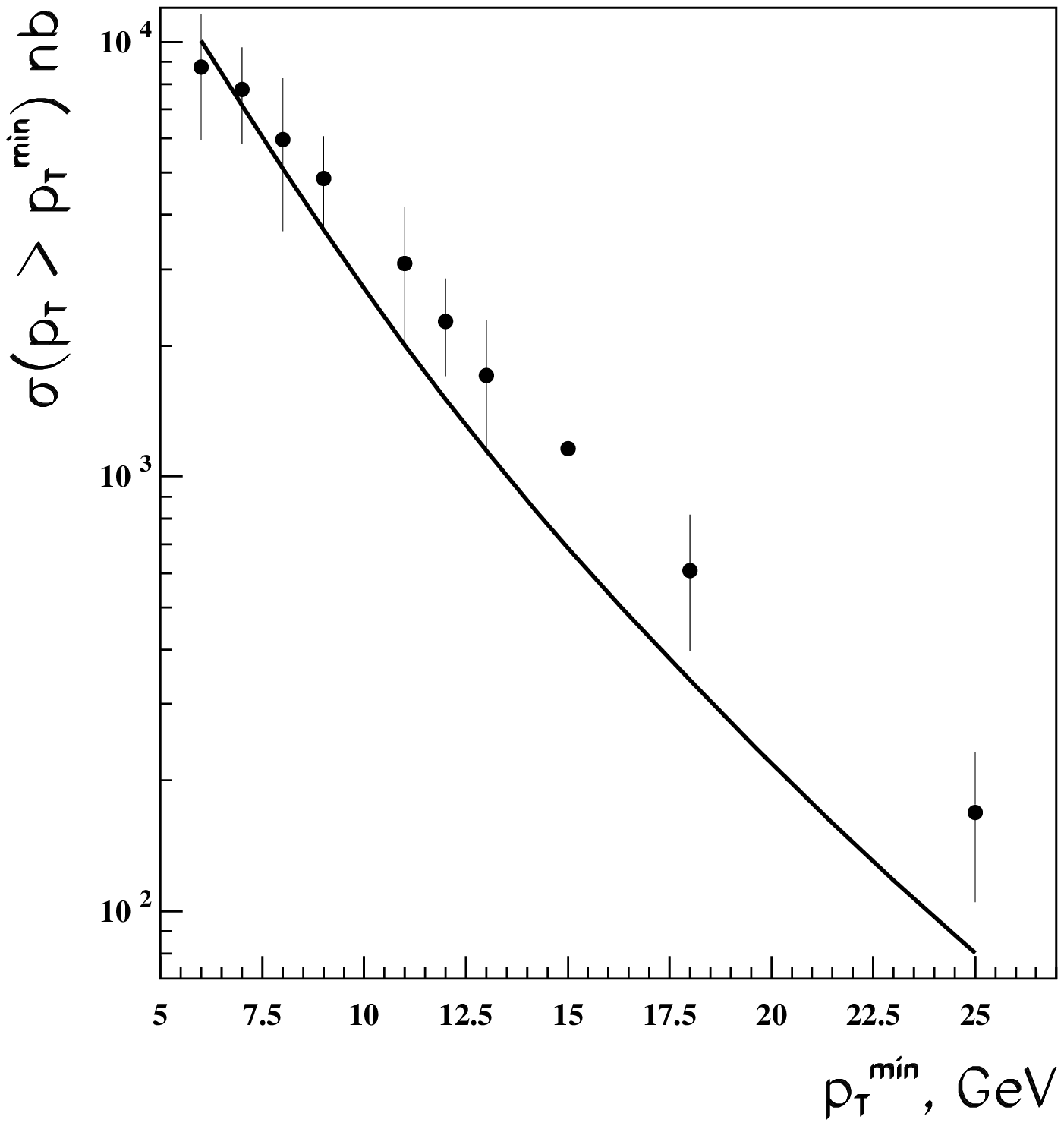}
\vskip -0.5 cm
\caption{\footnotesize
The calculated results for $b$ quark production
cross section $\sigma(p_T > p_T^{\mathrm min})$
at $\sqrt{s} = 1.8$~TeV and $|y|<1$.
The experimental data are taken from~\cite{Abbott:1999se}}
\label{bmin}
\end{figure}

The beauty production at the lower energy is also satisfactorily
reproduced as is shown in Fig.~\ref{bmin}, where the cross section of
$b$ quark production for $p_T > p_T^{\mathrm min}$ and $|y|<1$
is presented as a function of $p_T$.

\section{Conclusion}

We have demonstrated that the $k_T$ factorization method admits
reformulation in a manner making
it similar to the conventional parton model.
Likewise in the parton model
the sub process amplitude is written here for on shell
partons.
The integral over the transverse momenta of the incoming
off shell gluons, which is the main ingredient of
the $k_T$ factorization, turns here into the integral
over the orientation of the plane, where the incoming parton momenta lie.
The effect of this integrals, that
recovers the substantial part of the NLO parton model,
rapidly grows with the energy.
Their relative weight in the total cross section at
$\sqrt s = 27$~GeV is about 40\%, the rest comes from
the transverse momenta $|q_T^2|<Q_0^2$ corresponding
to the conventional LO parton model. At the same time
the LO parton model gives only 10\% at $\sqrt s=7$~TeV,
so that the $k_T$ factorization contribution dominates
at this energy.

Assumptions made to modify the $k_T$ factorization formalism
seem to be rather natural and not too restrictive.
The numeric
calculations confirm that they are sufficient for quite
reasonable description of the data.

As a rule our solid curves are slightly above the experimental
points. Most probably it is the consequence of the fact
that the produced heavy quarks can fragment
both into mesons and baryons
whereas the experiment data are presented mainly
for the secondary mesons.
The outcome meson/baryon ratio is close to 1/3
for the $c$ quark production in accordance with
a simple quark combinatorics,
(of course we can not say that 1/3 agrees with the
data better than, say, 1/5)
while this value
is clearly overestimated for the $b$ quark case.

To summarize, a large variety of experimental data
have been reproduced at least qualitatively with
two parameters, that is the mass
of the heavy quark and the factorization scale.
As the masses can not be varied in a wide ranges
it leaves only one actual parameter,
which allows nevertheless to obtain a reasonable
description of the experimental data
both for charm and beauty production
at different energies.

The authors are grateful to M.G.~Ryskin for helpful discussion.

\end{document}